\documentclass[12pt,a4paper]{article}

\usepackage{amsfonts,amssymb,amsmath,bm,amsbsy,subcaption}
\usepackage{amsthm}
\usepackage{graphicx}
\usepackage{caption}
\usepackage{subcaption}
\usepackage{enumerate}
\usepackage[authoryear]{natbib}
\usepackage[unicode=true,pdfusetitle,
 bookmarks=true,bookmarksnumbered=false,bookmarksopen=false,
 breaklinks=false,pdfborder={0 0 1},colorlinks=true]{hyperref}
\hypersetup{linkcolor=red, citecolor=blue}
\usepackage{xcolor}

\bibpunct{(}{)}{,}{a}{,}{,}

\newtheorem{theorem}{Theorem}
\newtheorem{assumption}{Assumption}
\newtheorem{proposition}{Proposition}

\newtheorem{lemma}{Lemma}

\newtheorem{remark}{Remark}

\renewcommand{\thesection}{\arabic{section}}
\renewcommand{\theequation}{\arabic{section}.\arabic{equation}}

\def\R{{\mathbb{R}}}
\newcommand{\E}{E}
\newcommand{\I}{\operatorname{I}}
\newcommand{\T}{\operatorname{T}}
\newcommand{\B}{\operatorname{B}}
\newcommand{\A}{\operatorname{A}}

\renewcommand{\P}{\operatorname{P}}
\newcommand{\1}{\mathbf{1}}
\newcommand{\M}{\operatorname{M}}
\newcommand{\sign}{\operatorname{sign}}
\newcommand{\as}{\quad \mbox{ a.s.}}
\makeatletter
\newcommand{\labeltext}[2]{%
  \@bsphack
  \csname phantomsection\endcsname 
  \def\@currentlabel{#1}{\label{#2}}%
  \@esphack
}
\makeatother

\title{One-step smoothing splines instrumental regression \thanks{Jad Beyhum undertook most of this work while employed by CREST, ENSAI (Rennes). He is grateful for support from the Research Fund KU Leuven under grant  STG/23/014. Elia Lapenta acknowledges funding from the French National Research Agency (ANR) under grant  ANR-23-CE26-0008-01. Pascal Lavergne acknowledges funding from the French National Research Agency (ANR) under the Investments for
the Future program (Investissements d'Avenir, grant
ANR-17-EURE-0010).\newline The authors thank Jean-Pierre Florens, Xavier D’Haultfoeuille, Ben Deaner, Frank Windmeijer, Daniel Wilhelm, Aureo de Paula, Kaspar Wuthrich, Christoph Breunig, Joachim Freyberger, Elie Tamer, seminar and conference participants at the University of Bonn, the 2023     Encounters in Econometric Theory at Nuffield College, Oxford University, the 2023 Bristol Econometrics Study Group conference, and the 34th (EC)$^2$ conference in Manchester. The authors are also grateful to two referees of \textit{The Econometrics Journal} along with the co-Editor Dennis Kristensen.} }

\author{Jad~Beyhum \\ Department of Economics, KU Leuven
\and  Elia~Lapenta \\ ENSAE-CREST \and Pascal~Lavergne \\ Toulouse School of Economics}

\def\AmSTeX{$\cal A$\kern-.1667em\lower.5ex\hbox{$\cal M$}\kern-.125em
    $\cal S$-\TeX}
\def\BibTeX{{\rm B\kern-.05em{\sc i\kern-.025em b}\kern-.08em
    T\kern-.1667em\lower.7ex\hbox{E}\kern-.125emX}}

\begin{document}

\maketitle

\begin{abstract}
We extend nonparametric regression smoothing splines to a context
where there is endogeneity and instrumental variables are available.
Unlike popular existing estimators, the resulting estimator is
one-step and relies on a unique regularization parameter. We derive
rates of the convergence for the estimator and its first
derivative, {\color{black}which are uniform in the support of the endogenous variable}.  We also address the issue of imposing monotonicity in
estimation and extend the approach to a partly linear model.
Simulations confirm the good performances of our estimator compared to
two-step procedures. Our method yields economically sensible results
when used to estimate Engel curves.

\noindent%

    \end{abstract}


\section{Introduction}
\setcounter{equation}{0}
    We consider the prototypical model
\begin{equation}
Y =   g_{0} (Z) +  \varepsilon \qquad \E [ \varepsilon |  W
] = 0 \, ,
\label{Model}
\end{equation}
where $Y \in \R$ is the dependent variable, $Z \in \R$ is the
endogenous continuous explanatory variable, and $W\in \R^{p}$ are the
instrumental variables (IVs).  The goal is to estimate
nonparametrically $g_0$, the causal effect of the variable $Z$ on $Y$,
using $W$ to account for endogeneity.  If we assumed linear
relationships, we could use the two-stage least squares estimator: in
a first stage, one obtains the linear projection of $Z$ given $W$,
then in a second stage one linearly regresses $Y$ onto the previously
estimated linear projection.

Considering a nonparametric function
$g_0$ allows estimating the causal relationship of $Y$ and $Z$ in a
more flexible manner.
Existing nonparametric estimators of $g_0$ typically rely on
two steps.
\cite{newey2003instrumental} develop a nonparametric equivalent to the
two-stage least squares estimator: they use  linear-in-parameter
series expansions of $\E [Y | W ]$ and
$\E[ g(Z) |W ]$ in a generalized method of moments
framework, see also \cite{ai2003efficient}, \cite{hall2005nonparametric},
\cite{blundell2007semi}, \cite{Johannes2011}, \cite{horowitz2014adaptive} for other
series-based methods. Alternatively, \cite{Florens2003}, \cite{hall2005nonparametric},
\cite{darolles2011nonparametric}, and \cite{gagliardini2012tikhonov}
rely on kernel methods to estimate the unknown conditional
expectations.  As is well-known, backing up a nonparametric estimate
of $g_0$ is an ill-posed inverse problem. Hence, one needs some kind
of regularization, such as hard thresholding, see
\cite{horowitz2011applied}, \cite{chen2012estimation}, Tikhonov or ridge-type
regularization, see \cite{newey2003instrumental}, \cite{darolles2011nonparametric},
\cite{florens2011identification}, \cite{gagliardini2012tikhonov}, \cite{singh2019kernel},
or a Landweber-type iterative method, see \cite{Johannes2013}, \cite{Dunker2014}.
A general exposition of some of these methods is given by \cite{carrasco2007linear}.
A recent machine learning literature considers solving a saddle
point problem that is dual to a generalized method of moments criterion.
Here one first maximizes an objective function with respect to
a function of the instruments $W$, then one minimizes with respect to
a function of $Z$ to obtain $g_0$, see, e.g., \cite{bennett2019deep}.
If the set of functions upon which one optimizes is
large, then one has in addition to introduce some penalization in the
optimization problem, see \cite{dikkala2020minimax}, \cite{Liao2020}.
\cite{muandet2020dual} consider a related but different saddle point problem.

We here develop a smoothing splines instrumental regression estimator for
$g_0$ that fully avoids nonparametric first stage
estimation. {\color{black} For a random sample $
\left(Y_i,Z_i,W_i\right), i=1, \ldots n$,  it is obtained  as
\begin{equation*}
    \widehat g(z)  =  \widehat a_{0} + \widehat a_{1} z + \frac{1}{12}
    \sum_{i=1}^{n}{ \widehat \delta_{i}
|z-Z_{i}|^{3}}\, ,
\end{equation*}
where the coefficients $\widehat a_0$, $\widehat a_1$, and $\widehat
\delta_i$ $(\text{with }i=1,\ldots,n)$ have a closed-form expression,
see Section \ref{sec:closed} for details. Hence, due to its spline
nature, our estimator is computationally simple and should be easily
appropriated by practitioners. Besides its simplicity,} our estimator
has several characteristics that should make it appealing for
empirical work. First, our approach is particularly attractive because
it is one-step. {\color{black} A key benefit of the one-step nature of our estimator
is that it depends upon one regularization parameter only.  In
existing two-step methods, each stage relies on a particular choice of
a smoothing or regularization parameter, whose fine-tuning may be
difficult in practice, while affecting the final results.
Two-step procedures typically lead to further difficulties, besides choosing
one tuning parameter at each step: one needs to estimate in a first
step an object that may be more complex than the final object of
interest, while first-stage estimation typically affects the
second-stage small sample and asymptotic properties.}  In some
methods, a third parameter is introduced to deal with the ill-posed
nature of the inverse problem.  To choose our unique regularization
parameter, we devise a practical cross-validation method that yields
good performances in simulations. {\color{black} {\color{black} Second,}
as we show,} our estimator is a natural generalization of the popular
smoothing splines estimator, see
\cite{Wahba1990}, \cite{green1993nonparametric}.  The appeal of splines lies
in their simplicity together with their excellent approximations of
smooth functions, see \cite{schumaker_2007}. Splines-based methods have
been extensively studied, see, e.g., \cite{Hall2005}, \cite{Li2008}, \cite{Claeskens2009},
\cite{Schwarz2016}, and have been found to have excellent performances in
practice. {\color{black} Third,   as an additional advantage, one obtains straightforward
estimators of derivatives, that are   typically of practical interest.}

We also propose some extensions to our method. First, we show how to impose
monotonicity constraints by relying on a method proposed by
\cite{Hall2001}.  The constrained estimator is simple to implement in
practice. Second, to illustrate the
versatility of our method, we extend our results to a partly linear
model. When used for estimating Engel curves, our smoothing splines
estimator and its monotone constrained version yield comparable results, that
are reasonable from an economic viewpoint.

The paper is organized as follows. In Section \ref{sec.est}, we detail
the {\color{black} construction} of our estimator.  We exhibit a global quantity that
accounts for all the information contained in Model (\ref{Model}), and
that is minimized by the true function $g_0$. We then consider an
empirical equivalent, and we set up a minimization problem penalized
by a roughness measure of the function to regularize the solution.  We
show that our estimator extends smoothing splines to the
instrumental variables context, and we give a closed form formula for
its computation.  The asymptotic properties of our estimator are
analyzed in Section \ref{sec: Asymptotic analysis}, where uniform
rates of convergences are derived for the function itself and its
derivative. {\color{black} We do not however address the issue of rate
optimality, see \cite{chen2011rate}.}
{\color{black} Section \ref{sec.cv} focuses on the choice of
the regularization via cross-validation.}
Section \ref{sec: Estimation under shape constraints}
deals with estimation under monotonicity constraints.  In Section
\ref{sec.impl}, we report selected simulation results, where
our estimator exhibits excellent finite sample performance compared to
some existing two-step methods, and we illustrate our method for Engel
curves estimation.  Finally, we extend our estimator to the partly
linear model in Section
\ref{sec.pl}. Concluding remarks are given in Section \ref{sec7}.
Details and supplementary results of simulations, as well as the proof of subsidiary results are included in the online appendix.

\section{Our estimator}\label{sec.est}
\setcounter{theorem}{0}
\setcounter{equation}{0}

\subsection{General formulation}
We assume that $g_{0}$ belong to some space of functions
$\mathcal{G}$ on which identification holds, that is,
\begin{equation}
\E [ Y -  g (Z)  |  W ]= 0 \as
\Rightarrow  g  = g_{0} \as
\label{ident}
\end{equation}
{\color{black} For a discussion of this condition called {\em
completeness}, see \cite{newey2003instrumental}, \cite{DHaultfoeuille2011}, and \cite{Freyberger2017}.}
{When $Z$ is continuous, as we assume here, $W$
should typically have   at least one continuous component for
 completeness to hold. Some of the instruments, however, could be
 discrete, and this will not affect further our exposition and reasoning.}

Instead of dealing directly with (\ref{ident}), as done by most
previous work, we consider an equivalent formulation that does not
require estimating a conditional expectation given the instruments
$W$.
{By the results of \cite{Bierens1982},
\begin{equation}
\E [ Y -  g (Z)  |  W ]= 0  \as \Leftrightarrow
\E [ (Y - g (Z)) \exp(\mathbf{i} W^\top t)] = 0 \quad \forall t\in\R^p
\, .
\label{eq:bierens}
\end{equation}
Consider now
\begin{equation}
\label{populationprogram}
g_0 =\arg\min_{g\in \mathcal{G}} M(g),
\qquad
M(g) =
\int \left| \E [ \left( Y - g (Z) \right) \exp(\mathbf{i}W^\top t) ] \right|^2 \, d\mu(t)
\, ,
\end{equation}
where $\mu$ is a symmetric probability measure {\color{black} with
support $\R^p$}. Then it is straightforward to see that $M(g)\geq 0$ for
all $g \in \mathcal{ G}$, and that under (\ref{ident})
\[
M(g)= 0  \Leftrightarrow g = g_{0} \as
\]
With a random sample $\left\{ \left(Y_i,Z_i,W_i\right), i=1,
\ldots n \right\}$ at hand, a natural estimator of  $M(g)$ is
\begin{align}
M_{n}(g)   & =
\int \left| \frac{1}{n} \sum_{i=1}^{n}{ \left( Y_i - g (Z_i) \right)
\exp(\mathbf{i}W_i^\top t) } \right|^2 \, d\mu(t)
\nonumber
\\
& =
\frac{1}{n^2}
\sum_{1\leq i, j \leq n }
\left( Y_i - g (Z_i) \right) \left( Y_j - g (Z_j) \right) \omega(W_i - W_j)
\, ,
\label{Mn}
\end{align}
{where}
\[
\omega(z) = \int_{\R^{p}} \exp(i t'z) \, d\mu(t) = \int_{\R^{p}} \cos(t'z) \, d\mu(t)
\, ,
\]
due to the symmetry of $\mu$: $\omega$ is (up to a constant) the
Fourier transform of the density of $\mu$.  The above formulation as a
V-statistic will be used in practice for computational purposes.

The condition for $\mu$ to have support $\R^{p}$ translates into the
restriction that $\omega$ should have a strictly positive Fourier
transform almost everywhere.  Examples include products of triangular,
normal, logistic, see \citet[Section 23.3]{JKB95}, Student, including
Cauchy, see \cite{DK2002}, or Laplace densities.  To achieve scale
invariance, we recommend, as in \cite{Bierens1982}, to scale the
exogenous instruments by a measure of dispersion, such as their
empirical standard deviation.  Note that the function $\omega$ is
\textit{not} similar to a typical kernel used in nonparametric
estimation, as there is no smoothing parameter entering $\omega$,
which is thus a fixed function that does not vary with the sample
size. Hence, our estimation procedure introduces no smoothing on the
instruments.}

If $W$ has bounded support, results from \cite{Bierens1982} yield that
 the equivalence (\ref{eq:bierens}) holds when $t$ is restricted to
 lie in a (arbitrary) neighborhood of $0$ in $\R^p$. Hence, $\mu$ can
 be taken as any symmetric probability measure that contains $0$ in
 the interior of its support.  As noted by \cite{Bierens1982}, there
 is no loss of generality assuming a bounded support, as his
 equivalence result equally applies to a one-to-one transformation of
 $W$, which can be chosen with bounded image.

{\color{black} Our statistic accounts for an infinity of moment conditions,
as stated in (\ref{eq:bierens}). It is different from
a generalized method of moments  (GMM) criterion that accounts for an
increasing  but finite number of moment
conditions, as used by e.g. \cite{ai2003efficient} and \cite{chen2012estimation}.
One can
nonetheless relate our estimator to GMM, in a way similar to
\cite{Carrasco2000}. As can be checked, under suitable conditions, the
function $\omega$ is a positive definite kernel, that is for any
sequence $z_1,
\ldots z_n$, and all real numbers $a_1, \ldots a_n$,
\[
\sum_{1\leq i, j \leq n }{ \omega(z_i-z_j) a_i a_j} = \int_{}^{}{\left|
\sum_{i=1}^{n}{a_i \exp(i z_i^\top t )}\right|^2 \, d\mu(t) }\geq 0, \quad
\, .
\]
Using Mercer's theorem, one can thus write
$\omega(t_i - t_j) = \sum_{k = 1}^{\infty}{\lambda_k e_k(t_i)
e_k(t_j)} $ for some infinite sequence of nonnegative eigenvalues $\lambda_k$, and an
orthonormal basis of functions $\left\{ e_k, k = 1, \ldots
\right\}$ in the space of square integrable functions. Hence,
\[
M_n(g)  = \sum_{k = 1}^{\infty} \lambda_k
\left[ \frac{1}{n} \sum_{i=1}^{n}  \left( Y_i - g (Z_i)\right) e_k(Z_i)  \right]^2
\, ,
\]
which is a GMM criterion that accounts for the countable sequence of moment conditions $\E
\left[\left(Y - g(Z) \right) e_k(W) \right] = 0$ for which $\lambda_k
\neq 0$. As shown in the Supplementary Appendix, if all $W_i$'s are
different, then all eigenvalues $\lambda_k$ are strictly positive, and
our criterion accounts for an infinitely countable number of moment
conditions.}

Minimizing $M_n(g)$ would lead to interpolation. We regularize the
problem by assuming some smoothness for the function $g$. We assume
that $Z$ has compact support, say $[0,1]$ without loss of generality,
and that $\mathcal{ G}$ is the space of differentiable functions on
$[0,1]$ with absolutely continuous first
derivative.\footnote{{\color{black}For the derivation of our asymptotic
results, when $Z$ has a compact support, say $[a,b]$, we can always
normalize it on $[0,1]$ by taking the affine transformation $z\mapsto
(z-a)/(b-a)$ and obtain the new variable $\widetilde{Z}:=(Z-a)/(b-a)$
supported on $[0,1]$. Then, we can consider the regression function
$z\mapsto g_0((b-a)z+ a)$ of $\widetilde Z$ defined on $[0,1]$, and our
theoretical analysis remains the same.}} That is, if $g
\in \mathcal{G}$, there is an integrable function $g''$ such that
$\int_0^z g''(t) \, dt = g'(z) - g'(0)$.  We then estimate $g_{0}$ as
a minimizer of a penalized version of $M_n(g)$ on $\mathcal{G}$.
{\color{black} Specifically,
\begin{equation}
\label{estimator}
\widehat{g} \in  \arg\min_{g\in \mathcal{G}}
S_n(g)\, , \qquad
S_n(g)  = M_n(g)  + \lambda \int_0^1 |g''(z)|^2\, dz
\, ,
\end{equation}
where $\lambda>0$ is a regularization parameter.}

A recent approach we became aware of when preparing this paper is the
``kernel maximum moment loss" approach proposed by
   \cite{zhang2023instrumental}.  While it does not smooth on the
   instruments, it assumes that the regression of interest belongs to
   a Reproducing Kernel Hilbert Space (RKHS), and solves a minimization
   problem by penalizing by the norm on such a space. The estimator
   thus depends on the chosen RKHS.  Differently, we assume that the
   regression of interest belongs to a space of smooth functions, and
   we penalize by the integral of the squared second derivative of the
   regression, which is a very intuitive measure of roughness, but not
   a RKHS norm.

\subsection{Closed-form solution}

\label{sec:closed}

We here show that (\ref{estimator}) has a unique solution, a natural
cubic spline, that we characterize in Proposition \ref{prop:1}
below.  We begin with some definitions.
{\color{black} For $a < t_1 < \ldots  < t_n < b$, a function $g$
on  $[a,b]$  is a cubic spline if two conditions are satisfied: on each of the
intervals $(a, t_1), \ldots (t_n, b)$, $g$ is a cubic polynomial;
the polynomial pieces fit together at each $t_i$ in such a way that $g$
and its first and second derivatives are continuous.
The points $t_i$ are called knots.}
A cubic spline on $[a,b]$ is said to be a {\em natural cubic spline} if its second and
third derivatives are zero at $a$ and $b$.  Without loss of
generality, we consider hereafter that $[a,b] = [0,1]$.

Given any values $(g_i, Z_i), \ i=1, \ldots, n$, $n \geq 2$, there is a unique
interpolating natural cubic spline, that is, a natural cubic spline
$g$ with knots {\color{black} at observations} $Z_i$ such that $g(Z_i) =
g_i, i = 1,\ldots n$. For details, see
e.g. \cite{green1993nonparametric}. A key result for our analysis is
the following.
\begin{theorem}[{\citet[Th. 2.3]{green1993nonparametric}}]
\label{ncsthm}
Suppose $n \geq 2$, $0 \leq Z_1 < \cdots < Z_n \leq 1$, and let $g$ be
the interpolating natural cubic spline with values $g_i$ at knots
$Z_i$, $i = 1\ldots n$.  Let $\widetilde{g}$ be any function in
$\mathcal{G}$ for which $\widetilde{g}(Z_i) = g_i, \ i=1, \ldots,
n$. Then
\begin{equation}
\label{ncs}
\int_0^1 |\widetilde{g}''(t)|^2\, dt \geq
\int_0^1 |g''(t)|^2\, dt
\, .
\end{equation}
with equality only if $\widetilde{g} = g$.
\end{theorem}
This result allows us to restrict our attention to natural cubic
splines, when studying the potential minimizers of {\color{black} $S_n(g)$}.
Indeed, suppose $\widetilde{g}$ is any function in ${\cal G}$ that is
not a natural cubic spline with knots at $Z_i$.  Let $g$ be the
natural cubic spline interpolant to the values
$\widetilde{g}(Z_i)$. Then $M_n\left({g}\right) =
M_n\left(\widetilde{g}\right)$.  Because of the above optimality
property of the natural cubic spline interpolant, (\ref{ncs}) holds
with strict inequality, and thus $ S_n(\widetilde{g}) > S_n(g)$. This
means that, unless $\widetilde{g}$ itself is a natural cubic spline
with knots at $Z_i$, we can find a natural cubic spline with knots at
$Z_i$ that attains a smaller value of $S_n(g)$. It follows at once
that a minimizer $g$ of $S_n(g)$, if it exists, must be a natural
cubic spline.  It is key to notice that we have not forced $g$ to be a
natural cubic spline. This arises as a mathematical consequence of the
choice of the roughness penalty. Now, as detailed below, we only need
to minimize $S_n(g)$ over a finite-dimensional class of functions.

{Assuming the $Z_i$'s are all different, which happens with
probability one for a continuous $Z$, a  natural cubic spline with
knots at $Z_i$ can be written as}
\begin{equation}
g(z)  =  a_{0} + a_{1} z + \frac{1}{12} \sum_{i=1}^{n}{ \delta_{i}
|z-Z_{i}|^{3}}\, ,
\quad
\sum_{i=1}^{n}{ \delta_{i}} = \sum_{i=1}^{n}{\delta_{i} Z_{i}} = 0
\, ,
\label{eq:gdecomp}
\end{equation}
{see \citet[Section 7.3]{green1993nonparametric}}. The function $g$ is
 uniquely defined by the coefficients
$a_0$, $a_1$, and $\delta_i, i = 1, \ldots n$, or equivalently by its
value at the knots, see Proposition \ref{prop:1}'s proof for details.

It will be useful for what follows to use matrix
notations.  Let
\[
\bm{Z}= \left(
\begin{array}{cc}
1 & Z_{1}
\\
\vdots & \vdots
\\
1 & Z_{n}
\end{array}
\right)
\, ,
\]
$\bm{E}(Z_1, \ldots Z_n) = \bm{E} = \left[ \frac{1}{12} |Z_{i}-Z_{j}|^{3} , i,j=1,\ldots n\right]$, and $\bm{g} =
 \left(g(Z_1), \ldots g(Z_n)\right)^T$.
Then $\bm{g} = \bm{Z a} + \bm{E \delta}$
with constraints $\bm{Z^T \delta} = 0$.  Also, one can check that
\[
\int  g'' (z)^{2} \, dz  =
\bm{\delta^T {E} \delta}
\, ,
\]
see \citet[Section 7.3]{green1993nonparametric}.
Let $\bm{Y}$ be the vector $\left(Y_1, \ldots Y_n\right)^T$, then
\begin{equation}
 M_{n}(g) + \lambda \int  \left(g'' (z)\right)^{2} \, dz
 =
\left( \bm{Y - Z a - E \delta}  \right)^T \bm{\Omega} \left( \bm{Y} - \bm{Z a} -
\bm{E \delta}  \right)
+ \lambda \bm{\delta^T {E} \delta}
\, ,
\label{eq:min}
\end{equation}
where $\bm{\Omega}$ is the matrix with generic element $n^{-2} \omega(W_i-W_j)$.
Hence, we want to minimize  a quadratic function in parameters
under the constraints   $\bm{Z^T \delta} = 0$. This yields a unique
solution under the usual requirements. The following proposition gives
a precise characterization.
\begin{proposition} \label{prop:1}
{For any $\lambda >0$, if all $Z_i$'s are different} and  all $W_i$'s are
different, the solution to (\ref{estimator}) exists, is unique, and is
a natural cubic spline $\widehat{g}$ characterized by
\begin{align}
\left[
\begin{array}{cc}
\widetilde{\bm{E}} & \bm{Z}
\\
\bm{Z^T} & \bm{0}
\end{array}
\right]
\left(
\begin{array}{c}
\bm{\widehat{\delta}}
\\
\bm{\widehat{a}}
\end{array}
\right)
& =
\left(
\begin{array}{c}
\bm{Y}
\\
\bm{0}
\end{array}
\right)
\, ,
\qquad
\widetilde{\bm{E}} = \bm{E} + \lambda \bm{\Omega}^{-1}
\, .
\label{estimformula}
\end{align}
{Moreover, the values at the knots are}
\[
\widehat{\bm{g}}  =
\left[
\bm{P} +
\bm{E} \widetilde{\bm{E}}^{-1} \left(\bm{I}-\bm{P}\right)
\right]
\bm{Y}
\, ,
\]
where  $\bm{P}= \bm{Z} \left(\bm{Z}^T \widetilde{\bm{E}}^{-1} \bm{Z}
\right)^{-1} \bm{Z}^T \widetilde{\bm{E}}^{-1}$.
\end{proposition}

Our estimator is obtained directly by solving the linear system of equations
(\ref{estimformula}). It does not necessitate estimation of other
nonparametric quantities, and relies on only one regularization
parameter $\lambda$. It also directly provides an estimator of the
first derivative of $g$ as
\begin{equation}
\label{eq: ghat prime}
\widehat{g}'(z)  =  \widehat{a}_{1} + \frac{1}{4} \sum_{i=1}^{n}{
\widehat{\delta}_{i} \sign(z-Z_i) (z-Z_{i})^{2}}
\, ,
\qquad
\sign(u) = \1(u \geq 0) - \1(u < 0)
\, .
\end{equation}

There are alternative ways to (\ref{eq:gdecomp}) for expressing a
natural cubic spline.  We focus on this formulation as it does not
rely on a particular support of $Z$, nor on the fact that the knots
$Z_i$ are arranged in increasing order. In particular, the closed-form
expression in Proposition \ref{prop:1} is valid regardless of the
support of $Z$ and therefore it can be used without first transforming
$Z$ into $(0,1)$.  We also found this formulation to be convenient for
practical implementation. For large samples, where the above formula
may not be computationally efficient, one can adapt to our context the
Reinsch algorithm, see \cite{green1993nonparametric}.

{\color{black} We note that it would be possible to generalize our
estimator (\ref{estimator}), say by penalizing by the integral of the square of the
third derivative, so as to obtain a natural quartic spline, see
\cite{Wahba1990}. Such a generalization would entail more
technicalities without changing the wide picture. In practice,
computational methods for storage of and computation based on natural
cubic splines are well developed, see
e.g. \cite{green1993nonparametric} and references therein. Therefore,
in our work, we focus on this case.}

\section{Asymptotic analysis}\label{sec: Asymptotic analysis}
\setcounter{theorem}{0}
\setcounter{equation}{0}

The formal study of our estimator is based on a reformulation of
$M(g)$ in (\ref{populationprogram}). Let $\mathcal{D}^2$ be the set of
twice weakly differentiable functions. Consider
\[
\mathcal{G}=\left\{ g:[0,1] \to \R, g \in \mathcal{D}^2 :
  \int_{0}^1 |g''(t)|^2 \, dt < \infty \right\},
\,
\qquad
\mathcal{H}=\left\{ h \in  \mathcal{G}:
h(0) = h'(0) = 0 \right\},
\,
\]
and the inner product $\left<h_1,h_2\right>_{\mathcal{H}}=\int_{0}^1
h_1''(z)\,h_2''(z)\, dz$ on $\mathcal{H}$.\footnote{\color{black} The
considered inner product on $\mathcal{H}$ is standard in the spline
literature, see \cite{Wahba1990}. Other work where penalization on
derivatives is used also considers inner products implying
derivatives, see e.g. \cite{florens2011identification} or
\cite{florens2018nonparametric}. Note there is a one-to-one
correspondence between a function   $h \in \mathcal{H}$  and its
second derivative.}  Each $g\in
\mathcal{G}$ can be uniquely written as $g(z)=(1,z)\beta+h(z)$, where
$\beta=(g(0), g'(0))\in \mathbb{R}^2$, $h(z)=g(z)-g(0)-g'(0)z$,
$h\in\mathcal{H}$.  Denote by $L^2_\mu $ the space of complex
functions $l$ from $\mathbb{R}^q$ onto $\mathbb{C}$ such that
\[
\|l\|^{2}_\mu  = \int_{}^{}{|l(t)|^2 \, d\mu(t)} < \infty
\, .
\]
Consider the  operators $\A:\mathcal{H}\mapsto
L^2_\mu $ and $\B:\R^2\mapsto L^2_\mu $ such that
\begin{align}\label{eq: definitions of of A and B}
     \A h=\E[h(Z)\exp(\mathbf{i}W^T\cdot)] \quad \text{and} \quad \B
     \beta=\E[(1,Z)\beta \exp(\mathbf{i}W^T\cdot)]\, .
\end{align}
The minimization problem  (\ref{populationprogram}) identifying $g_0$ can be expressed as
\begin{equation}
\label{eq: Partly Linear population program}
   \min_{\beta,h} \|\E[Y \exp(\mathbf{i}W^T\cdot)]-\B\beta - \A h \|_\mu^2
\end{equation}
for $(\beta,h)\in\mathbb{R}^2\times\mathcal{H}$.  The above quantity
reaches its minimum zero at $(\beta_0,h_0)$, with
$g_0(z)=(1,z)\beta_0+h_0(z)$. A key advantage of this formulation for
theoretical analysis is that using orthogonal projections, we can
profile (\ref{eq: Partly Linear population program}) to first
determine ${h}_0$, then ${\beta}_0$ as a function of ${h}_0$. In our
proofs, we will also consider the  penalized empirical
counterpart of (\ref{eq: Partly Linear population program}) and use a
similar profiling method to obtain $\widehat g(z)$.
The following assumption ensures that $\E[Y\exp(\mathbf{i}W^T\cdot)]\in
L^2_{\mu}$, and that $\A$ and $\B$ are valued in
$L^2_{\mu}$.
\begin{assumption}\label{ass:square integrability} (a) $\E[Y^2]<\infty$; (b)
    $Z$ has a bounded density $f_Z$ on $[0,1]$;
(c) $\mu$ is  a symmetric probability measure with support $\R^p$;
(d) $\int\int \E[\exp(\mathbf{i}W^\top  t)f_{Z}(z)]^2\mu(t) dz \,
dt<\infty$; (e) $W$
has at least one  continuous component.
\end{assumption}
Our assumptions on the support of $Z$ is without much loss of
generality, since  we can always use a one-to-one
transformation that maps $Z$ into $[0,1]$.
We  then formalize the completeness assumption, under which the
problem (\ref{eq: Partly Linear population program}) admits a
unique solution $(\beta_0,h_0)$.
\begin{assumption}\label{ass:completeness}
$g_0$ belongs to $\mathcal{G}$ and
the mapping $g\in \mathcal{G} \mapsto \E[g(Z)|W=\cdot]$ is injective.
\end{assumption}
We now introduce a {\em source condition}, which is common in the
literature on inverse problems.  While it is not
needed to establish the consistency of $\widehat g$ and its first
derivative, it is necessary to obtain convergence rates.
\begin{assumption}
\label{ass:source condition}
Let $\M$ be the orthogonal projection onto the orthogonal of the span
of $\B$, and let $\T = \M \A$.  Let $(\sigma_j,\varphi_j,\psi_j)_j$ be
the singular system of $\T$, where $(\varphi_j)_j$ is a sequence of
orthonormal elements in $\mathcal{H}$, $(\psi_j)_j$ is a sequence of
orthonormal elements in $L^2_\mu$, and $(\sigma_j)_j$ is
a sequence of strictly positive values in $\mathbb{R}$. Then there
exists $\gamma>0$ such that
$$
\sum_j
\sigma_j^{-2\gamma}|\left<h_0,\varphi_j\right>_{\mathcal{H}}|^2<\infty
\,.
$$
\end{assumption}
\textcolor{black}{ As noted in \citet[page 1550]{darolles2011nonparametric}, the $\gamma$ in our source condition  is related to (i) the smoothness of the function $g_0$, as measured by the rate of decay of the Fourier coefficients $(\left<g_0,\varphi_j\right>_{\mathcal{H}})_j$; and (ii) the degree of ill-posedness of the inverse problem, as measured by the rate of decay of the singular values $(\sigma_j)_j$. Again following \cite{darolles2011nonparametric}, the singular values $(\sigma_j)_j$ are related to the link between the endogenous regressor and the instruments. For example, in the extreme case where the endogenous regressor and the instruments are independent, there is at least one null singular value.  }

\begin{theorem}\label{th.cvrates}
    Under Assumptions \ref{ass:square integrability},
\ref{ass:completeness},  if     $\lambda\rightarrow
    0$ and $n \lambda \rightarrow \infty$, then
    \begin{align*}
        \sup_{z\in[0,1]}|\widehat g(z)- g_0(z)|=o_p(1)\, \text{ and }
    &\sup_{z\in[0,1]}|\widehat g'(z)- g'_0(z)|=o_p(1)\, .
    \end{align*}
    If moreover Assumption \ref{ass:source condition} holds, then
    \begin{equation*}
    \sup_{z\in[0,1]}|\widehat g(z)- g_0(z)| \text{ and }
    \sup_{z\in[0,1]}|\widehat g'(z)-
    g'_0(z)|
    \text{ are both }
    O_p\left(\frac{1}{\sqrt{n \lambda}}+\lambda^{\frac{\gamma
    \wedge 2}{2}}\right)\, .
    \end{equation*}
\end{theorem}

We obtain consistency of our estimator and its derivative under mild
assumptions, that only involve a standard condition on the
regularization parameter $\lambda$. \color{black}The role of the condition $n\lambda\to \infty$ is to ensure that the variance term vanishes, while the restriction $\lambda\to 0$ guarantees that the bias goes to 0.\color{black}
\ {\color{black}By contrast, in} two-step estimation methods that
smooth over the instruments, one has to ensure that first-step
estimation is consistent, and one typically needs conditions that
relate the different smoothing parameters, see
e.g. \cite{ai2003efficient}, \cite{chen2012estimation}. In some
instances, consistency may further necessitate regularization
parameters, see \cite{chen2012estimation}.
A general discussion can be found in \cite{carrasco2007linear}.

Turning now to our rates of convergence, we do not claim that these are
sharp. {\color{black} As noted by a referee, our
rates cannot be minimax, in the sense of \cite{chen2011rate}, since we
are not bounding away from infinity the norm of  functions in ${\cal
G}$. \cite{hall2005nonparametric} and
\cite{chen2018optimal} also obtain estimators with minimax rates.
It is, however, unclear how to compare our results to these optimal rates,
which are in $L^2$ norm and hold under a different set of
assumptions. In the Supplementary Appendix, we study the $L^2$ norm of
our estimator and its derivatives under a set of assumptions that
resemble the ones of \cite{chen2011rate}, namely, we rely on Hilbert
scales and a link condition.

By contrast to previous results in this literature, our rates depend
upon only one smoothing parameter.  Our assumption, though, assumes
that the problem is mildly ill-posed, while some previous work also
considers the case of severely ill-posed inverse problems. As apparent from our proofs,} the rate
$1/\sqrt{n\lambda}$ corresponds to a standard
deviation term, while the second rate $\lambda^{\frac{\gamma \wedge
2}{2}}$ corresponds to a bias term.  If $\lambda$ is chosen to balance
these two rates, we obtain the convergence rate $n^{- \frac{\gamma
\wedge 2}{2\left(1+ \gamma \wedge 2\right)}}$.  For $\gamma = 2$ or 1,
this respectively yields $n^{-1/3}$ and $n^{-1/4}$.

{\color{black}

\section{Choice of regularization parameter}\label{sec.cv}

 We propose to choose our single penalty parameter $\lambda$ via
cross-validation, where the criterion to be minimized is akin to
$M_n(g)$.  The simplest procedure is based on random sample splitting,
where an estimator $\widehat{g}_\lambda$ is obtained from the first
fold of size $n_1 =
\lfloor n/2\rfloor$ (the largest integer smaller than $n/2$), and the second  fold of size $n_2= n -
n_1$ is used for validation. The criterion
\begin{equation}
M_{n_2}(\widehat{g}_\lambda)
 = \int \left| \frac{1}{n_2} \sum_{i=n_1 +1}^{n}{ \left( Y_i - \widehat{g}_\lambda (Z_i) \right)
\exp(\mathbf{i}W_i^\top t) } \right|^2 \, d\mu(t)
\nonumber
\label{Mnlambda}
\end{equation}
is thus computed and minimized over a  fixed set
 $\Lambda=[\underline{\lambda}_n,\overline{\lambda}_n]\subset
 \mathbb{R}_+$ of  penalty parameters, where
 $(\underline{\lambda}_n)_n$ and $(\overline{\lambda}_n)_n$ are two
 sequences  such that $\overline{\lambda}_n\rightarrow
 0$ and  $n \underline{\lambda}_n\rightarrow \infty$.
A more elaborate procedure consists in letting each fold play the role
of the training fold in turn. Specifically, we compute $\widehat{g}_{k,
\lambda}$ for each fold $k = 1,2$, and we create the cross-validated
criterion
\[
\int \left|
\frac{1}{n}
\left( \sum_{i=1}^{n_1}{ \left( Y_i - \widehat{g}_{2,\lambda} (Z_i) \right)
\exp(\mathbf{i}W_i^\top t) } +
 \sum_{i=n_1 + 1}^{n}{ \left( Y_i -
\widehat{g}_{1,\lambda} (Z_i) \right) \exp(\mathbf{i}W_i^\top t) }
\right)
\right|^2 \, d\mu(t)
\, .
\]
This two-fold cross-validation method can easily be extended to
k-fold cross-validation, or repeated k-fold cross-validation.
In empirical implementations, we used the
above two-fold cross-validation method with good results.

For such a data-driven procedure, we here provide some theoretical
rationale, similar to the ones put forward by, e.g.,
\cite{darolles2011nonparametric}. Technical details
can be found in the supplementary material.  To simplify exposition,
let us consider sample splitting (similar arguments could be applied
to k-fold cross-validation schemes). Then
\begin{align*}
\frac{1}{n_2} \sum_{i=n_1+1}^{n}{ \left( Y_i - \widehat{g}_\lambda (Z_i) \right)
\exp(\mathbf{i}W_i^\top t) }&  =
\frac{1}{n_2} \sum_{i=n_1+1}^{n}{ \left( Y_i - {g}_0 (Z_i) \right)
\exp(\mathbf{i}W_i^\top t) }
\\ \mbox{}  & +
\frac{1}{n_2} \sum_{i=n_1+1}^{n}{ \left( {g}_0 (Z_i) - \widehat{g}_\lambda (Z_i) \right)
\exp(\mathbf{i}W_i^\top t) }
\, .
\end{align*}
 Under suitable assumptions, using simple arguments, one can show that the first term is
 $O_p (n^{-1/2})$ in $\|\cdot\|_\mu$ norm.  For the second term, one can show that
\[
\left| \frac{1}{n_2} \sum_{i=n_1+1}^{n}{ \left( {g}_0 (Z_i) - \widehat{g}_\lambda (Z_i) \right)
\exp(\mathbf{i}W_i^\top t)} \right|
\leq
\sup_z |g_0(z) - \widehat{g}_\lambda(z)|
= O_P \left( \frac{1}{\sqrt{n\lambda}} + \lambda^{\frac{\gamma \wedge 2}{2}}\right)
\]
uniformly over $\Lambda$. Hence,
$M_{n_2}(\widehat{g}_\lambda) =  O_P \left(\frac{1}{{n\lambda}} +
\lambda^{\gamma \wedge 2}\right)$ uniformly over  $\Lambda$, which is  the
same order as $\sup_z |g_0(z) - \widehat{g}_\lambda(z)|^2$.

Minimizing our criterion with respect to $\lambda$ should thus give a
penalty parameter that yields the appropriate balance between bias and
variance. This argument is similar to the one put forward in previous
works on data-driven choices of  penalty parameters in Tikhonov
nonparametric instrumental regressions, see
\cite{feve2010practice,feve2014non,Carrasco2014}, and also
\cite{engl1996regularization} for a  Tikhonov  estimator with a known
operator.
These works similarly show that their
criterion used to select the penalty
parameter is an $O_P$ of the same order as the rate of convergence of
their estimator, and argue that optimizing their criterion with respect to the regularization parameter should be
roughly equivalent to optimizing the convergence rate of their estimator.
A more detailed analysis would focus on the asymptotic properties of
the estimator that uses the above data-driven regularization parameter, but this is
outside the scope of the current paper.}

{
\section{Estimation under monotonicity}
\setcounter{theorem}{0}
\setcounter{equation}{0}
\label{sec: Estimation under shape constraints}

In some instances, we may expect the function of interest $g_0$ to be
monotonic.
If $g_0$ is the Engel curve that relates the proportion of consumer expenditure on a
good as a function of total expenditure, we typically expect this
function to be increasing for a “normal'' good and decreasing for an
“inferior'' good. Accounting for monotonicity  in
estimation is expected to improve accuracy in small and moderate
samples, {\color{black}see
\cite{mammen1999smoothing}} and
\cite{chetverikov2017nonparametric}.

To implement such a monotonicity constraint into estimation, we note
that since our smoothing splines estimator is linear, the derivative
estimator (\ref{eq: ghat prime}) is linear as well.  Let us express it in
matrix form.  
We can write
$
\bm{g'} = \bm{O a} + \bm{D \delta}
$,
where
$\bm{g'} = \left(g'(Z_1), \ldots g'(Z_n)\right)^T$,
 $\bm{D} = \left[ \frac{1}{4} \sign(Z_i - Z_j) |Z_{i}-Z_{j}|^{2} ,
 i,j=1,\ldots n\right]$, and
\[
\bm{O} =
\left(
\begin{array}{cc}
0 & 1
\\
\vdots & \vdots
\\
0 & 1
\end{array}
\right)
\, .
\]
From Proposition \ref{prop:1},
\begin{equation}
\label{eq:coefs}
\begin{pmatrix} \bm{\widehat{\delta}}\\\bm{\widehat a}
    \end{pmatrix}=
    \bm{S}
    \begin{pmatrix}\bm{Y}\\  \bm{0}
    \end{pmatrix}
    \, ,
    \quad
\bm{S}=\left[
\begin{array}{cc}
\widetilde{\bm{E}} & \bm{Z}
\\
\bm{Z^T} & \bm{0}
\end{array}
\right]^{-1}
\, ,
\quad \mbox{\rm so } \quad
\bm{g'} =
\left( \bm{D}, \bm{O} \right)
\bm{S}
\begin{pmatrix} \bm{Y} \\ \bm{0} \end{pmatrix}
\, .
\end{equation}
We rely on a method proposed by \cite{Hall2001}, that is based on the
same linear estimator but reweights the observations $Y_i$ to impose
monotonicity at observations points.  It adjusts the unconstrained
estimator by tilting the empirical distribution to make the least
possible change, in the sense of a distance measure, subject to
imposing the constraint of monotonicity at observation points.
Specifically, if $g_0$ is assumed to be monotonically increasing, we
consider the constrained optimization program
\begin{align}
\label{eq: constrained optimization problem}
\min_{p_1,\ldots,p_n} &n-\sum_{i=1}^n (n p_i)^{1/2}
\\
\text{ subject to } \quad
\sum_{i=1}^n p_i=1\,&,\quad p_i\geq 0\text{ for all }i=1,\ldots,n\quad
,\quad
\left( \bm{D}, \bm{O} \right)
\bm{S}
 \begin{pmatrix} \bm{p} \circ \bm{Y} \\
\bm{0}
\end{pmatrix} \geq 0
\, , \nonumber
\end{align}
where $\bm{p} \circ \bm{Y} = (p_1 Y_1,\ldots, p_n Y_n)^T$ is the
Hadamard product between vectors.
If $g_0$ was assumed to
be monotonically decreasing, we would modify
the last inequalities.
\cite{Hall2001} considered more general optimization problems based on
a family of Cressie-Read divergences, but we focus on the above
program for convenience. It is strictly convex, so it admits a unique
solution $\bm{p^*}$, and it is computationally fast to solve. The
final estimator  ${\widehat{g}^*}$ is the  natural cubic spline with coefficients
$\bm{a}^*$ and $\bm{\delta}^*$ defined as in (\ref{eq:coefs}), with
$\bm{p^*} \circ \bm{Y}$ in place of  $\bm{Y}$.

{\color{black} To obtain  the asymptotic properties of our constrained
smoothing splines estimator, we state the following supplementary assumption.}
\begin{assumption}\label{ass:monotonicity}
There exists $\eta>0$ such that $g_0'(z)\geq \eta$ for all $z\in[0,1]$.
\end{assumption}
{\color{black} The results  from Theorem
\ref{th.cvrates} then directly apply to our constrained estimator.}
Indeed, as $\widehat{g}'$ is uniformly consistent,
the constraint in the optimization problem (\ref{eq: constrained
optimization problem}) becomes asymptotically irrelevant from
Assumption \ref{ass:monotonicity}. Accordingly, $\widehat g^*=\widehat
g$ with probability approaching one, and our results readily follow.
While the monotonicity constraints become asymptotically irrelevant,
they can matter in finite samples, as shown by
\cite{chetverikov2017nonparametric} and illustrated by our empirical
results.}

\section{Numerical results}
\setcounter{equation}{0}
\setcounter{theorem}{0}
\label{sec.impl}

\subsection{Small sample behavior}
\label{sec.num}
We used a DGP in line with Equation (\ref{Model}), where
\begin{align*}
    \varepsilon=\frac{aV+\eta}{\sqrt{1+a^2}}\, ,\,&
    a=\sqrt{\frac{\rho_{\varepsilon V}^2}{1-\rho_{\varepsilon V}^2}}\,
    ,\\
    Z=\frac{\beta W + V}{\sqrt{1+\beta^2}}\,
    ,\,&\,\beta=\sqrt{\frac{\rho_{WZ}^2}{1-\rho_{WZ}^2}} \, ,
\end{align*}
and $(W,V,\eta)$ are independent standard Gaussians.
This yields standard Gaussian marginal distributions for $\varepsilon$
and $Z$ whatever the values of the parameters. The correlation
$\rho_{\varepsilon V}$ measures the level of endogeneity of $Z$. The
correlation $\rho_{WZ}$ measures instead the strength of the
instrument $W$.

We implemented our smoothing splines estimator with $\omega$
equal to the density of a Laplace distribution with mean zero and
variance 1.  The choice of the penalty parameter $\lambda$ was based
on two-fold cross-validation, as detailed {\color{black}in Section \ref{sec.cv}}.  We considered
 $\lambda$ within the grid  $\{p/(1-p),p=10^{-5}+k*(0.7-10^{-5})/399,k=0,\dots,399\}$.

\color{black}

We compared our estimator to two existing methods for which a
data-driven procedure has been proposed for the choice of smoothing
{or}  regularization parameters. We considered first \textcolor{black}{the Tikhonov estimator penalized by the Sobolev norm of degree $2$ of \cite{gagliardini2012tikhonov}}, hereafter
referred as Tikhonov. We also considered the series estimator of \citet{horowitz2014adaptive} based on a basis of Legendre
polynomials. The implementation details of both methods are given in the online supplement, together with supplementary results.

We first considered two functional forms for $g_0$, each normalized to
have unit variance: a quadratic function $g_{0,1}(z) =
{z^2}/{\sqrt{2}}$, and a non-polynomial function $g_{0,2}(z) =
\sqrt{3\sqrt{3}}\,z\,\exp(-z^2/2)$.  We ran 2000 Monte Carlo
simulations with sample sizes $n=200$ and $400$. We consider three
couples of values for $(\rho_{\varepsilon V},\rho_{WZ})$: (a)
$(0.5,0.9)$, a setting with low endogeneity and a strong instrument,
(b) $(0.8,0.9)$, corresponding to high endogeneity and a strong
instrument, (c) $(0.8,0.7)$, a more complex setting with high
endogeneity level but a weaker instrument. To evaluate the gains of imposing monotonicity, we then considered a
third function $g_{0,3}(z) = (\sqrt(10/3) \log(|z-1|+1) \sign(z-1) -
0.6 z+ 2 z^3)/8$.  The regularization parameter $\lambda$ was chosen
before the monotonizing step, and we used the {\tt R} package CVXR
to solve (\ref{eq: constrained optimization problem}), see \cite{Fu}.

\begin{table}[!ht]
\caption{Simulation results}
\begin{center}
\resizebox{0.9\textwidth}{!}{\begin{tabular}{rl|ccc|ccc|cccc}
  \hline
   & &\multicolumn{3}{c}{$g_{0,1}$} & \multicolumn{3}{c}{$g_{0,2}$} & \multicolumn{4}{c}{$g_{0,3}$}\\
   \hline
   $n$   &  & Sm. & Tikh. & Ser. & Sm.  & Tikh. & Ser. & Cons.  & Sm.  & Tikh. & Ser. \\
\hline
  & &\multicolumn{10}{c}{$\rho_{ZW}=0.9\,,\, \rho_{\varepsilon V}=0.5$} \\ \hline

200 & Bias$^2$ & 0.000 & 0.029 & 0.092 & 0.001 & 0.005 & 0.005 & 0.003 & 0.000 & 0.022 & 0.103\\
200 & Var & 0.069 & 0.170 & 0.060 & 0.074 & 0.072 & 0.073 & 0.041 & 0.077 & 0.219 & 0.115\\
200 & MSE & 0.069 & 0.199 & 0.152 & 0.075 & 0.076 & 0.078 & 0.044 & 0.077 & 0.241 & 0.218\\
\hline
400 & Bias$^2$ & 0.000 & 0.028 & 0.094 & 0.001 & 0.002 & 0.005 & 0.001 & 0.000 & 0.043 & 0.140\\
400 & Var & 0.052 & 0.152 & 0.031 & 0.053 & 0.050 & 0.025 & 0.026 & 0.057 & 0.188 & 0.044\\
400 & MSE & 0.052 & 0.180 & 0.125 & 0.054 & 0.052 & 0.030 & 0.027 & 0.057 & 0.231 & 0.184\\
\hline
  & &\multicolumn{10}{c}{$\rho_{ZW}=0.9\,,\, \rho_{\varepsilon V}=0.8$} \\
  \hline
200 & Bias$^2$ & 0.001 & 0.031 & 0.092 & 0.001 & 0.004 & 0.005 & 0.003 & 0.000 & 0.021 & 0.106\\
200 & Var & 0.066 & 0.169 & 0.060 & 0.072 & 0.075 & 0.074 & 0.039 & 0.073 & 0.205 & 0.106\\
200 & MSE & 0.067 & 0.200 & 0.152 & 0.072 & 0.079 & 0.079 & 0.043 & 0.073 & 0.227 & 0.212\\
\hline
400 & Bias$^2$ & 0.000 & 0.029 & 0.094 & 0.000 & 0.002 & 0.005 & 0.001 & 0.000 & 0.044 & 0.141\\
400 & Var & 0.049 & 0.151 & 0.030 & 0.051 & 0.050 & 0.026 & 0.025 & 0.054 & 0.179 & 0.042\\
400 & MSE & 0.050 & 0.179 & 0.125 & 0.052 & 0.052 & 0.030 & 0.026 & 0.054 & 0.223 & 0.183\\
\hline
  & &\multicolumn{10}{c}{$\rho_{ZW}=0.7\,,\, \rho_{\varepsilon V}=0.8$} \\ \hline
200 & Bias$^2$ & 0.009 & 0.052 & 0.133 & 0.004 & 0.018 & 0.158 & 0.024 & 0.012 & 0.034 & 0.189\\
200 & Var & 0.091 & 0.247 & 0.130 & 0.120 & 0.129 & 0.053 & 0.057 & 0.102 & 0.235 & 0.050\\
200 & MSE & 0.099 & 0.299 & 0.262 & 0.124 & 0.146 & 0.211 & 0.081 & 0.114 & 0.269 & 0.239\\
\hline
400 & Bias$^2$ & 0.003 & 0.053 & 0.139 & 0.004 & 0.011 & 0.157 & 0.010 & 0.002 & 0.024 & 0.190\\
400 & Var & 0.069 & 0.202 & 0.053 & 0.087 & 0.097 & 0.024 & 0.041 & 0.086 & 0.235 & 0.022\\
400 & MSE & 0.073 & 0.255 & 0.192 & 0.090 & 0.107 & 0.180 & 0.051 & 0.088 & 0.259 & 0.212\\
\hline
\end{tabular}}
\end{center}
\label{table: simulation results}
\footnotesize
\renewcommand{\baselineskip}{11pt}
\textbf{Note:} Average over a grid of 100 equidistant points on $[-2,2]$ and 2000 Monte Carlo replications of the squared bias (Bias$^2$), the variance (Var), and the Mean Squared Error (MSE) for the constrained smoothing splines estimator (Cons.), the smoothing splines estimator (Sm.),  the Tikhonov estimator (Tikh.), and the series estimator (Ser.).
\end{table}

Table \ref{table: simulation results} reports our results.
{\color{black} The series estimator is severely biased in all cases, but
for the non-polynomial $g_{0,2}$ and strong instruments, while our estimator is
always almost unbiased. The Tikhonov estimator mostly lies in between,
but with large differences depending on the functional form of
$g_0$. In terms of variance, the series estimator does slightly better than
smoothing splines in most cases, that itself does better than
Tikhonov, but for the non-polynomial $g_{0,2}$.  Smoothing splines performs best
in terms of MSE in almost all cases. Exceptions are cases
corresponding to the non-polynomial $g_{0,2}$ with $n=400$ and strong
instruments, where the series estimator is close to unbiased. Overall,
the severity of endogeneity does not affect much the estimators'
performances.  When the instrument's strength decreases, our smoothing
splines estimator appears to perform best.  Finally, imposing
monotonicity to our smoothing spline estimator does not affect much
bias, but yields a substantial decrease in variance, as
expected. Depending on the particular setup, it can be more than
halved.}

\subsection{Empirical application}
We applied the smoothing spline estimator to the estimation of Engel curves, which relate
 the proportion of spending on a given good as a
 function of total expenditures. We used the “Engel95'' dataset \citep[]{engel95},  from the {\tt R} package {\tt np}, see \cite{hayfield2008}. This dataset  is a random sample from the 1995 British Family Expenditure Survey and contains data for 1655
 households of married couples for which the head-of-household is
 employed and between 25 and 55 years old. We focused on the subsample
 of 628 households with no kids.  We report results for two Engel
 curves, pertaining to the expenditure shares on {leisure and fuel}.
 Economic theory suggests that the Engel curve for leisure is
 increasing and the one for fuel is decreasing.  Following
 \cite{blundell2007semi}, we instrumented the  \textit{logarithm of total household's expenditure},
 which is likely endogenous, by the
 \textit{logarithm of total earnings before tax}.  We consider the
 four estimators used in our simulations, and implementation details
 remain the same.

\begin{figure}[!ht]
\centering
\begin{subfigure}{0.5\textwidth}
  \centering
  \includegraphics[width=\linewidth]{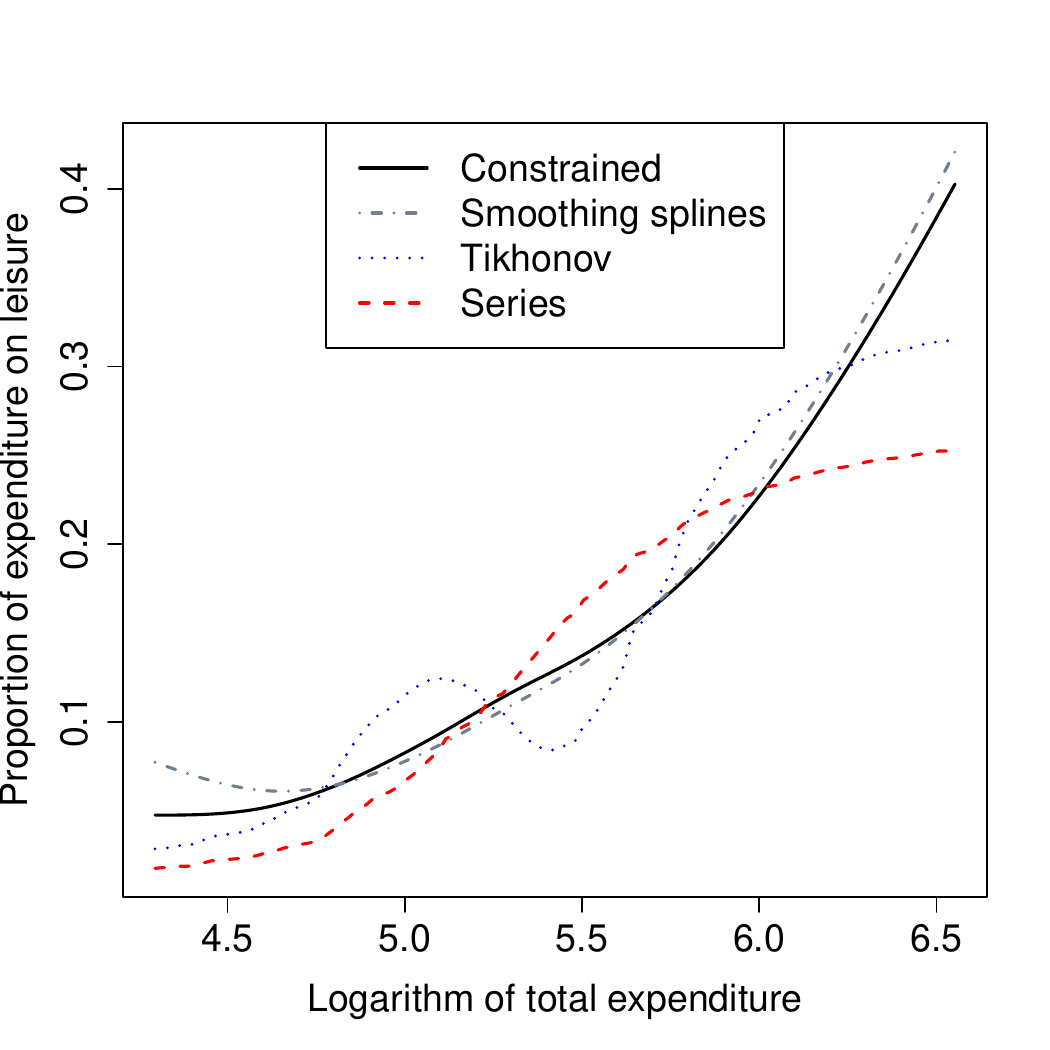}
  \caption{Leisure}
  \label{fig7:a}
\end{subfigure}%
\begin{subfigure}{0.5\textwidth}
  \centering
  \includegraphics[width=\linewidth]{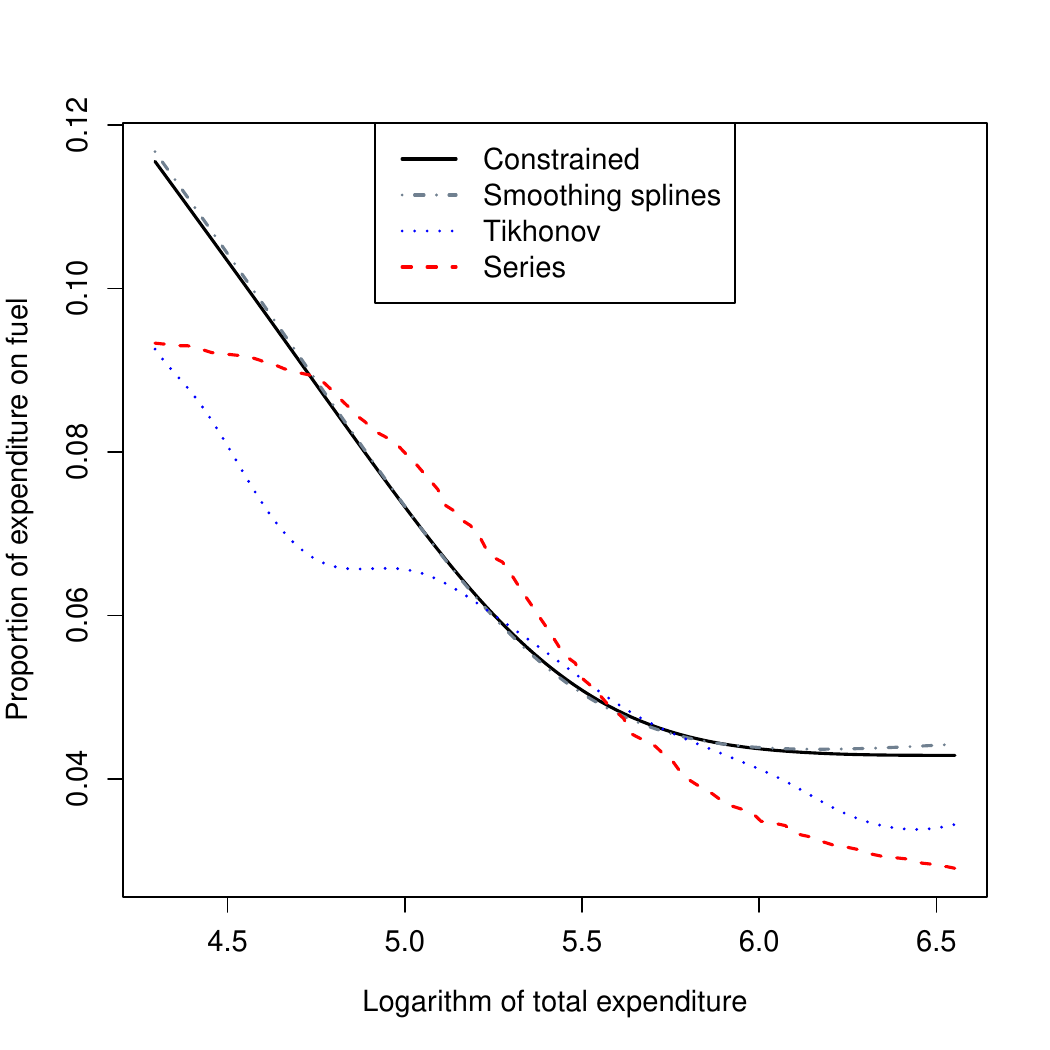}
  \caption{Fuel}
  \label{fig7:b}
\end{subfigure}
\caption{Estimated Engel curves.}
  \label{fig7}
\end{figure}

The  estimated nonparametric functions are reported in Figure  \ref{fig7}.
The Tikhonov estimate exhibits a non-monotonic and quite
 irregular behavior, while the series estimate is mainly monotonic and
 very regular.  Since our smoothing splines estimates are monotonic,
 but at the boundaries of the data, our constrained and unconstrained
 estimates are very close. Both are in line with the findings of
 \cite{blundell2007semi}.

\section{Extension to a partly linear model}
\setcounter{equation}{0}
\setcounter{theorem}{0}
\label{sec.pl}

We have proposed a generalization of regression smoothing splines to
the context where there is endogeneity and instrumental variables are
available.  While we detail our estimator and its properties in the
simple univariate context, a multivariate extension could be
considered. However, including more covariates in a fully
nonparametric way would submit us to the curse of dimensionality
typical of functional estimation.  Hence, we focus here on a partly
linear model, as considered by e.g. \cite{heckman1986Spline}, \cite{Robinson1988},
\cite{blundell2007semi}, \cite{Chen2009}, and
\cite{florens2012instrumental}.
This model provides a simple and
economical way to include additional controls.  We thus consider
\begin{equation}\label{eq: pl extension}
Y= X^T {\gamma}_0 + g_0(Z)+\varepsilon\quad  \E[\varepsilon|W]=0\,,
\end{equation}
where ${X} \in\mathbb{R}^q$ is a vector of exogenous covariates, whose
components are thus included in $W$, while, as earlier,
$Z\in\mathbb{R}$ is the endogenous variable.  The following condition
ensures  identification of $({\gamma}_0,g_0)$
in (\ref{eq: pl extension}).

\begin{assumption}\label{ass:completeness pl}
$({\gamma}_0,g_0)$ belongs to $\mathbb{R}^q\times \mathcal{G}$ and the
mapping $({\gamma}_0,g_0)\in \mathbb{R}^q\times \mathcal{G} \mapsto
\E[X^T{\gamma}+g(Z)|W]$ is injective.
\end{assumption}
Our identification assumption is similar to ones imposed in other
work, see, e.g., \cite{chen2012estimation} or \cite{florens2012instrumental}.
First, it excludes collinearity between the
components of $X$. Second, it requires that the distribution of $Z$
given $W$ must be complete so that the mapping $g\mapsto
\E[g(Z)|W]$ is injective. Third, it rules out the presence of an
intercept in $X$, since an intercept can always be absorbed
by the nonparametric function $g_0$. Fourth, it requires that
no function $g$ is such that $\E[g(Z)|W]$ is a linear function of the variables in $X$.
Under Assumption \ref{ass:completeness pl},
\begin{align*}
(\gamma_0,g_0)=\arg\min_{(\gamma,g)\in
\mathbb{R}^{q}\times\mathcal{G}}
\int_{ }|\E[(Y-X^T\gamma -
g(Z))\exp(\mathbf{i}W^T t)]|^2 d \mu(t)
\, .
\end{align*}
We proceed as in the benchmark model and estimate
$(\gamma_0,g_0)$ by minimizing the empirical counterpart of 
the above criterion penalized by a roughness measure of the nonparametric component, that is
\begin{align}\label{estimator pl extension}
& (\widehat \gamma,\widehat g)\in  \arg
\min_{(\gamma,g)\in\mathbb{R}^{q}\times\mathcal{G}}
M^{PL}_n(\gamma,g)+\lambda \int |g''(z)|^2 dz,\\
 \text{where }M_n^{PL}(\gamma,g)& =
 \int_{ }\left| \, \frac{1}{n}\sum_{i=1}^n
 (Y_i-X_i^T\gamma-g(Z_i))\exp(\mathbf{i}W^T t)\, \right|^2 d \mu(t)\, .
 \nonumber
\end{align}
The estimators $(\widehat \gamma,\widehat g)$ can be computed in the
same way as in the benchmark model. To show this, let $\bm{L}$ be the
$n\times (q+2)$ matrix whose row $i$ is
$(1,Z_i,X_i^T)$.
The following  is a direct extension of Proposition \ref{prop:1}
to the partly linear model.
\begin{proposition} \label{prop:1 pl extension}
 For any $\lambda >0$, if all $Z_i$'s and  all $W_i$'s are
different, and $\mathbf{L}$ is full-column rank, the solution to (\ref{estimator pl extension}) exists and is
unique. The estimator $\widehat{g}$ is a natural cubic  spline.
The coefficients  $\widehat{\bm{a}}  = \left(  \widehat{a}_0,
\widehat{a}_1, \widehat{\gamma}^T\right)^T$ and $\bm{\delta}$
are characterized by
\begin{align}
\left[
\begin{array}{cc}
\widetilde{\bm{E}} & \bm{L}
\\
\bm{L}^T & \bm{0}
\end{array}
\right]
\left(
\begin{array}{c}
\bm{\widehat{\delta}}
\\
\bm{\widehat{a}}
\end{array}
\right)
& =
\left(
\begin{array}{c}
\bm{Y}
\\
\bm{0}
\end{array}
\right)
\, ,
\qquad
\widetilde{\bm{E}} = \bm{E} + \lambda \bm{\Omega}^{-1}
\, .
\label{estimformula for pl}
\end{align}
\end{proposition}

We now focus on the consistency and the convergence rates of the
regression function. Let us define the operator $\operatorname{D}:\mathbb{R}^{q+2}\mapsto L^2_\mu$ such that
\begin{equation}\label{eq: definition of D}
    \operatorname{D} (\gamma,\beta)=\E[(X^T\bm{\gamma} + (1,Z)\beta)\exp(\mathbf{i}W^T\cdot)]\, .
\end{equation}
\begin{assumption}\label{ass: square integrability for pl}
    Assumption \ref{ass:square integrability} holds and $\E\|X\|^2<\infty$.
\end{assumption}
\begin{assumption}\label{ass:source condition for pl}
    Assumption \ref{ass:source condition} holds with
    $\operatorname{D}$ replacing $\operatorname{B}$ and
    $\operatorname{M}$ being the orthogonal projection operator onto
    the orthogonal of the span of $\operatorname{D}$.
\end{assumption}

The following  is a direct extension of Theorem \ref{th.cvrates}.
\begin{theorem}\label{th.cvrates pl}
    Under Assumptions \ref{ass:completeness pl} and \ref{ass: square
    integrability for pl}, if $\lambda\rightarrow 0$ and $n \lambda
    \rightarrow \infty$, then
        \begin{align*}
        \sup_{z\in[0,1]}|\widehat g(z)- g_0(z)|=o_p(1)\, \text{ and }
    &\sup_{z\in[0,1]}|\widehat g^{'}(z)- g'_0(z)|=o_p(1)\, .
    \end{align*}
    If moreover Assumption \ref{ass:source condition for pl} holds, then
    \begin{equation*}
    \sup_{z\in[0,1]}|\widehat g (z)- g_0(z)|
    \text{ and }
    \sup_{z\in[0,1]}|\widehat g^{'} (z)-
    g'_0(z)|
    \text{ are both }
    O_p\left(\frac{1}{\sqrt{n \lambda}}+\lambda^{\frac{\gamma
    \wedge 2}{2}}\right)\, .
    \end{equation*}
\end{theorem}

\color{black}

\section{Concluding remarks}\label{sec7}

Further extensions to our method could be considered, such as a
nonparametrically additive model, see \cite{Linton1995}, \color{black} a
more detailed theory for the choice of the regularization parameter, and
inference results on the nonparametric function of interest or some
functional of it.

Concerning the data-driven selection of $\lambda$,
\cite{breunig2016adaptive,breunig2020adaptive}, and \cite{chen2024adaptive} have
derived theoretical results for a Lepski-type selection procedure for
series nonparametric instrumental variables estimators without
penalization. An interesting avenue for research is to extend such
results to Tikhonov estimators such as ours. \cite{jansson2020towards} provides high-level conditions for the Lepski's method for regularized estimators. One could try to verify these conditions for our approach.

Regarding inference results, most of the literature has focused on
non-penalized estimators for nonparametric instrumental variable
regressions. Two exceptions are \cite{chen2015sieve}, which contains
inference results on functionals of a penalized series estimator,
and \cite{babii2020honest}, which develops honest and uniform
confidence bands for a Tikhonov estimator. The results in
\cite{chen2015sieve} need to be significantly adapted. Indeed, the latter
paper considers a finite but growing number of moments while we consider infinitely many moments. Moreover, the penalized sieve minimum distance
estimator studied \cite{chen2015sieve} combines two types of
regularizations: slowly growing finite-dimensional sieves and
penalization, while we only rely on the latter. Concerning the
approach of  \cite{babii2020honest}, we consider a different Tikhonov
estimator with other operators, so the proofs must also be adapted.
\color{black}

\bigskip
\bibliographystyle{apalike}
\bibliography{biblio_EJ}

\section*{Appendix A: Proof of Theorem 3.1}
\renewcommand{\theequation}{A.\arabic{equation}}
\renewcommand{\thesection}{A}
\setcounter{equation}{0}
\setcounter{theorem}{0}

\medskip

\textbf{Proof of Theorem \ref{th.cvrates}:}
We start by introducing some useful notations and results. Let
 $\mathcal{X}$ and $\mathcal{Y}$ be Hilbert spaces with corresponding
 inner products $\left<\cdot,\cdot\right>_\mathcal{X}$ and
 $\left<\cdot,\cdot\right>_\mathcal{Y}$, and consider a linear
 operator $\operatorname{D}:\mathcal{X}\mapsto \mathcal{Y}$.  The norm
 of $\operatorname{D}$ is
 $\|\operatorname{D}\|_{op}=\sup_{f\in\mathcal{X},\|f\|_{\mathcal{X}}=1}\|\operatorname{D}
 f\|_\mathcal{Y} $.  When $\|\operatorname{D}\|_{op}<\infty$,
 $\operatorname{D}$ is said to be bounded (or continuous), see
 \citet[Chapter 2]{kress1999linear}.  Let $\operatorname{D}^*$ be the
 adjoint of $\operatorname{D}$, defined as
 $\operatorname{D}^*:\mathcal{Y}\mapsto \mathcal{X}$ such that
 $\left<\operatorname{D}
 f,\psi\right>_{\mathcal{Y}}=\left<f,\operatorname{D}^*\psi,\right>_{\mathcal{X}}$
 for any $(f,\psi)\in\mathcal{X}\times \mathcal{Y}$.  When
 $\operatorname{D}$ is bounded, $\operatorname{D}^*$ always exists and
 $\|\operatorname{D}\|_{op}=\|\operatorname{D}^*\|_{op}$, see
 \citet[Theorem 4.9]{kress1999linear}.
In what follows, we will repeatedly use the
 following properties: (a) $\|\operatorname{D}f\|_{\mathcal{Y}}\leq
\|\operatorname{D}\|_{op}\|f\|_{\mathcal{X}}$ for any
$f\in\mathcal{X}$, and  (b) if $\operatorname{C}$ is another linear
operator, then $\|\operatorname{C}\operatorname{D}\|_{op}\leq
\|\operatorname{C}\|_{op}\|\operatorname{D}\|_{op}$, whenever the
composition $\operatorname{C}\operatorname{D}$ is well defined.

We divide the proof into several steps. In Step 1, we analyze the
minimization problem at the population level. In Step 2, we analyze
the problem at the sample level.  In Step 3, we bound the norm of
$\widehat{h} - h_0$.  In Step 4 and 5, we combine the results to first
establish uniform consistency of $\widehat g$ and its first
derivative, second to obtain uniform rates of convergence.
The proof relies on Lemmas S3.1 and S3.2, which are stated in the online supplement.

\textbf{Step 1}.
From Assumption  \ref{ass:completeness} and \cite{Bierens1982},
\[
g = 0 \Leftrightarrow
\E [   g (Z)  |  W ]= 0 \Leftrightarrow
\E [   g (Z)  \exp(\mathbf{i} W^\top t) ] = 0 \ \forall
t\in\R^p
\, .
\]
Hence,  the null space of the linear mapping $g\mapsto
\E[g(Z)\exp(\textbf{i}W^T\cdot)]$ only
contains the zero element, and  such a mapping is injective (one-to-one).
This implies that $\A h=\E[h(Z)\exp(\mathbf{i}W^T\cdot)]$ and $\B \beta=\E[(1,Z)\beta
\exp(\mathbf{i}W^T\cdot)]$ are also injective.

Each $g\in\mathcal{G}$ can be uniquely written as
$g(z)=(1,z)\beta+h(z)$, where $\beta=(g(0),g'(0))$,
$h(z)=g(z)-g(0)-g'(0)z$,  $h \in\mathcal{H}$.
Hence, the intersection of the ranges of the operators $\A$ and $\B$ is
the null function, since
$\A{h}=\B{\beta}$ iff $ (1, z) \beta - h(z) = 0$.

Consider the problem
\begin{equation}
\label{eq: partly Linear population program}
   \min_{\beta,h} \| r-\B\beta - \A h
   \|_\mu^2
   \, , \qquad r=\E[Y\exp(\textbf{i}W^T \cdot)]
   \, ,
\end{equation}
where $\|\cdot\|_\mu$ is the $L^{2}_\mu$ norm.
If $g_0(z)=(1,z)\beta_0+h_0(z)$, then $(\beta_0,h_0)$ is the unique
solution. We now obtain an explicit expression of $(\beta_0,h_0)$
solving (\ref{eq: partly Linear population program}).  Let $\P$ be the
orthogonal projection operator of functions in $L^2_\mu$ onto
$\mathcal{R}(\B)$ the range of $\B$.  Since $\B$ is defined on
$\mathbb{R}^2$, its range $\mathcal{R}(\B)$ is a linear finite
dimensional space. As linear finite dimensional spaces are complete,
see \citet[Theorem 2.4-2 ]{kreyszig1978introductory},
$\mathcal{R}(\B)$ is also linear and complete. By \citet[Theorem
1.26]{kress1999linear}, projection operators onto linear and complete spaces are well-defined, and so is $\P$.

We now show that  $\P$ writes as $\B(\B^*\B)^{-1}\B^*$, where $\B^*$ is the adjoint of $\B$.  As
previously noted, $\B$ is injective and its null space is
$\mathcal{N}(\B)=\{0\}$. Then  $\mathcal{N}(\B^*\B)=\mathcal{N}(\B) =
\{0\}$, $\B^*\B$ is injective, and $(\B^*\B)^{-1}$ exists.
As linear operators mapping $\mathbb{R}^2$ into $\mathbb{R}^2$ are
uniquely characterized by
second order matrices, see \citet[Section 2.9]{kreyszig1978introductory}, $\B^*\B$ is a second order
matrix, as well as its inverse. Hence, the operator
$\B(\B^*\B)^{-1}\B^*:L^2_\mu\mapsto
L^2_\mu $ is well-defined.
For any $f\in L^2_\mu$ and $\beta\in\mathbb{R}^2$,
$
    \left< f-\B(\B^*\B)^{-1}\B^*f,\B\beta\right>_\mu=\left<\B^*f-\B^*f,\beta\right>=0\, .
$
Hence, $f-\B(\B^*\B)^{-1}\B^*f\perp \mathcal{R}(\B)$, and $\B(\B^*\B)^{-1}\B^* f$ indeed represents the projection of $f$ onto $\mathcal{R}(\B)$, see  \citet[]{kress1999linear}. Therefore, $ \P =\B(\B^*\B)^{-1}\B^*$.

Let $\M = \operatorname{I}-\P$ be the orthogonal projection onto the orthogonal complement of $\mathcal{R}(\B)$. Then
$
r=\B \beta_0 +\A h_0 \Rightarrow \M r = \M \A h_0  = \T h_0
\, .
$
The operator $\T= \M \A$ is injective, since  the intersection of the
  ranges of  $\A$ and $\B$ is the null  function and $\A$  is
  injective. This yields
  \[
   h_0=\T^{-1}\M r\, , \qquad \beta_0=(\B^*\B)^{-1}\B^*(r-A h_0)\, .
\]
Consider now the penalized problem
\begin{equation}
\label{eq: partly linear integral equation penalized}
\min_{(\beta,h)\in\mathbb{R}^2 \times \mathcal{H}}
\| r - {\A}h - {\B}\beta\|^2_{\mu}+\lambda \|h\|^2_{\mathcal{H}}
\, .
\end{equation}
Let us profile with respect to $\beta$. For any fixed $h$,
\[
\min_{\beta\in \mathbb{R}^2 } \| r - {\A}h - {\B}\beta\|^2_{\mu}
 =
\| r - {\A}h - {\P}(r-\A h) \|^2_{\mu}
 =
\|\M r - {\T}h  \|^2_{\mu}
\, .
\]
We thus need to solve
\[
\min_{h\in\mathcal{H}}
\| \M r - {\T}h \|^2_{\mu}+\lambda \|h\|^2_{\mathcal{H}}
\, .
\]
From Lemma S3.2(a), ${\T}$ is compact, and thus
bounded. A direct application of
\citet[Theorem 16.1]{kress1999linear} ensures that the unique solution
$h_\lambda$ satisfies
$({\T}^*{\T}+\lambda \I) h_\lambda={\T}^*{\M} {r}$.
Now, for any $h$,
\[
\lambda \| h \|^{2}_{\mathcal{H}} \leq \lambda \| h \|^{2}_{\mathcal{H}} + \|\T h\|^{2}_\mu
=
\lambda \left<h,h\right>_{\mathcal{H}} + \left<h,\T^{*} \T h\right>_{\mathcal{H}}
=
\left<h, ({\T}^*{\T}+\lambda \I) h\right>
\, .
\]
Hence, $({\T}^*{\T}+\lambda \I)$ is strictly coercive and has a bounded
inverse by the Lax-Milgram
Theorem, see \citet[Theorem 13.26]{kress1999linear}. Therefore,
\begin{equation}
 h_\lambda= ({\T}^*{\T}+\lambda \I)^{-1} {\T}^*{\M} {r}
 \, .
 \label{eq:hlambda}
\end{equation}

\textbf{Step 2}. We study the minimization problem at the sample level
and we obtain sample counterparts of the population objects of Step 1.
Recall that  $\widehat g$ solves
\begin{equation}\label{eq: sample penalized program}
  \min_{g\in\mathcal{G}} \int_{ }\left|\frac{1}{n}\sum_{i=1}^n [Y_i -
  g((Z_i))]\exp(\textbf{i} W_i^T t)\right|^2 \mu(d  t)+\lambda\int_{0}^1 |g''(z)|^2 dz\, .
\end{equation}
By Proposition \ref{prop:1}, under Assumption \ref{ass:square
integrability}, the solution $\widehat{g}$ is unique with probability
1, and since each
$g\in\mathcal{G}$ writes uniquely as $g(z)=(1,z)\beta+h(z)$,
there is a unique $(\widehat \beta, \widehat h)$ such that
$\widehat{g}(z)=(1,z)\widehat{\beta}+\widehat{h}(z)$.  Define
\begin{align}
\widehat{\A}:\mathcal{H}\mapsto L^2_\mu\, , & \qquad
\widehat{\A}h =\frac{1}{n}\sum_{i=1}^n h(Z_i)    \exp(\textbf{i}W_i^T
\cdot)\, , \\
\widehat{\B}:\mathbb{R}^2\mapsto L^2_\mu\, , & \qquad
\widehat{\B}\beta =\frac{1}{n}\sum_{i=1}^n (1,Z_i)\beta
\exp(\textbf{i}W_i^T \cdot)\, ,
\end{align}
and $\widehat r =(1/n)\sum_{i=1}^n Y_i \exp(\textbf{i}W_i^T \cdot)$.
The optimization problem   (\ref{eq: sample penalized program}) is equivalent to
\begin{equation}\label{eq: sample penalized program with beta and h v2}
\min_{(\beta,h)\in\mathbb{R}^2 \times \mathcal{H}}
\|\widehat{r} - \widehat{\A}h - \widehat{\B}\beta\|^2_{\mu}+\lambda \|h\|^2_{\mathcal{H}}\, .
\end{equation}
We will  profile with respect to $\beta$, and to do so requires dealing
with the orthogonal projection onto the range of $\widehat{\B}$. Let
us proceed as in Step 1.
First,
\[
\|\widehat{\B}\beta\|^2_{\mu}
= 0 \Leftrightarrow
\int_{ }\left|\frac{1}{n}\sum_{i=1}^n (1,Z_i)\beta
\exp(\textbf{i}W_i^T t)\right|^2 \mu(dt)=0 \Leftrightarrow
\beta^T\bm{Z}^T\bm{\Omega}\bm{Z}\beta=0
\, .
\]
From Assumption \ref{ass:square integrability}(e), $W$ has at least
one continuous component, so that all $W_i$'s are different with
probability  1, and thus $\bm{\Omega}>0$ with probability
 1 from the proof of Proposition \ref{prop:1}.  Hence,
$\|\widehat{\B}\beta\|^2_{\mu} = 0$ iff $\bm{Z}\beta=0$.  As $\bm{Z}$
has full column rank with probability  1 from Assumption
\ref{ass:square integrability}, $\|\widehat{\B}\beta\|^2_{\mu} = 0$
iff $\beta=0$, and $\B$ is injective.  Let $\widehat{\P}$ be the
orthogonal projection onto the range of $\widehat{\B}$, which is well
defined and can be expressed as $\widehat{\P}
=\widehat{\B}(\widehat{\B}^*\widehat
\B)^{-1}\widehat{\B}^*$. Then,
$
\min_{\beta\in\mathbb{R}^2}\|\widehat r -
\widehat{\A}h-\widehat{\B}\beta\|^2_\mu
=
\|\widehat{r}-\widehat{\A}h-\widehat{\P}(\widehat{r}-\widehat{\A} h)\|_\mu=
\|\widehat{\M} \widehat r - \widehat{\T} h\|^2_\mu
\, ,
$
where    $\widehat{\M} = I-\widehat{\P}$ and
$\widehat{\T}=\widehat{\M} \widehat{\A}$.
We thus need to solve
\begin{align}\label{eq: sample profiled min program}
\min_{h\in\mathcal{H}}\|\widehat{\M} \widehat{r} -
\widehat{\T}h\|^2_\mu + \lambda \|h\|^2_\mathcal{H}\, .
\end{align}
From Lemma S3.2(e), $\widehat{\T}$ is compact, and thus
bounded. Thus, using a similar reasoning as in Step 1, the unique solution  is
\begin{align}\label{eq: expression of hhat}
  \widehat h = (\widehat{\T}^*\widehat{\T}+\lambda \I)^{-1}
  \widehat{\T}^*\widehat{\M} \widehat r
\, ,
\end{align}
which in turn yields
\begin{equation}
\label{eq: expression for betahat}
 \widehat
 \beta=(\widehat{\B}^*\widehat{\B})^{-1}\widehat{\B}^*(\widehat r -
 \widehat{\A}\widehat h)
 \, .
 \end{equation}

\textbf{Step 3}.
We now  prove that
\begin{equation}
\label{eq: rate for hhat depending on the regularization bias}
    \|\widehat h - h_0\|_\mathcal{H}=O_p\left( \frac{1}{\sqrt{n
    \lambda}}+\|h_{\lambda}-h_0\|_\mathcal{H}\right)
    \, .
\end{equation}

We consider  the  decomposition
$\widehat h - h_0=S_1+S_2+S_3+S_4+h_\lambda-h_0$, where
\begin{gather*}
    S_1=(\T^* \T + \lambda \I )^{-1}\T^* (\widehat{\M}\widehat r-
    \widehat{\T} h_0)\, , \quad
    S_2=(\T^* \T + \lambda \I )^{-1}(\widehat{\T}^*-
    \T^*)(\widehat{\M}\widehat r - \widehat{\T}h_0)\, ,\\
    S_3= \left[(\widehat{\T}^* \widehat{\T}+\lambda \I)^{-1} - (\T^* \T
    + \lambda \I )^{-1} \right]\widehat{\T}^* (\widehat{\M}\widehat r
    - \widehat{\T}h_0)\, ,  \quad
    S_4=(\widehat{\T}^* \widehat{\T}+\lambda \I)^{-1}\widehat{\T}^*
    \widehat{\T}h_0 - h_\lambda\, .
\end{gather*}
\textcolor{black}{Since $\lambda\rightarrow 0$ and $n \lambda\rightarrow \infty$, a direct application of Lemma S3.3(a) to (d) yields  $\|S_j\|_{\mathcal{H}}=O_P(1/\sqrt{n \lambda})$ for $j=1,2,3,4$. So, we obtain   (\ref{eq: rate for hhat depending on the
regularization bias}).}

\textbf{Step 4}.
We here show convergence of our estimators. Since
$\T$ is injective from Step 1 and  compact from Lemma
S3.2(a),  $\|({\T}^* \T +\lambda {\I})^{-1}
{\T}^*{\T} h - h \|_{\mathcal{H}} = o(1)$ for all $h$ whenever
$\lambda \rightarrow 0$, see \citet[Definition 15.5 and Theorem
15.23]{kress1999linear}. Hence $
\|h_\lambda-h_0\|_{\mathcal{H}}=o(1)$.
This and (\ref{eq: rate for hhat depending on the regularization bias}) yields $\|\widehat h - h_0\|_{\mathcal{H}}=o_p(1)$ if
in addition $n\lambda \rightarrow \infty$.

We now show that $ \|\widehat \beta -    \beta_0
\|=O_p\left(1/\sqrt{n} +\|\widehat   h-h_0\|_\mathcal{H}
\right)$. From (\ref{eq: expression for betahat}),
\begin{align*}
    \widehat{\beta}-\beta_0=&
    [(\widehat{\B}^*\widehat{\B})^{-1}- (\B^* \B)^{-1}]
    \widehat{\B}^*(\widehat r - \widehat{\A}\widehat h)
    + (\B^*\B)^{-1}
    [\widehat{\B}^*-\B^*](\widehat{r}-\widehat{\A}\widehat h) \\
    & + (\B^* \B)^{-1} \B^* (\widehat r - r)
     + (\B^* \B)^{-1} \B^* (\A-\widehat{\A})\widehat h
     - (\B^* \B)^{-1} \B^*\A(\widehat h - h_0)
\\
\Rightarrow \|  \widehat{\beta}-\beta_0 \|  \leq &
\|(\widehat{\B}^*\widehat{\B})^{-1}- (\B^* \B)^{-1}\|_{op} \,
\|\widehat{\B}^*\|_{op}\,
\left(\|\widehat r\|_{\mu}+\|\widehat \A\|_{op}\,\|\widehat
h\|_{\mathcal{H}}\right)
\\
& +
\|(\B^* \B)^{-1}\|_{op}\, \|\widehat{\B}^* - B^*\|_{op}\,
\left(\|\widehat r\|_{\mu}+\|\widehat \A\|_{op}\,\|\widehat
h\|_{\mathcal{H}} \right)
\\
& +
\|(\B^* \B)^{-1}\|_{op}\, \| B^*\|_{op}\,
\left(\|\widehat r - r \|_{\mu} + \|\widehat \A - \A \|_{op}
\,\|\widehat h\|_{\mathcal{H}}  + \| \A \|_{op}
\,\|\widehat h - h_0\|_{\mathcal{H}}\right)
\, .
\end{align*}
Lemma S3.1 ensures that
$\|(\widehat{\B}^*\widehat{\B})^{-1}- (\B^* \B)^{-1}\|_{op}$,
$\|\widehat{\B}^* - \B^*\|_{op} = \|\widehat{\B} - \B\|_{op}  $,
$\|\widehat r - r \|_{\mu}$, and $\|\widehat \A - \A \|_{op}$ all are
$O_p(n^{-1/2})$.
We have  $ \| B^*\|_{op} = \|\B\|_{op} <
\infty$, as $\B$ is a linear operator  with  finite dimensional
domain, see  \citet[Theorem
2.7-8]{kreyszig1978introductory}, and
$\|\B^* \B\|_{op} = \|\B\|_{op}^2$. Similarly, $\|(\B^* \B)^{-1}\|_{op}
<\infty$ as $\B$ is injective.
From Lemma S3.2(a), $A$ is compact and hence
bounded, and from Lemma S3.2(d)
$\|\widehat{\A}\|_{op}=O_p(1)$.
From a similar reasoning, $\|\widehat{\B}^*\|_{op}=O_p(1)$.
Also $\|\widehat h - h_0\|_{\mathcal{H}}=o_p(1)$ implies
$\|\widehat h \|_{\mathcal{H}}=O_p(1)$.
Combine these results to obtain that
$$
\|  \widehat{\beta}-\beta_0 \|  = O_p\left( n^{-1/2} +
\|\widehat h - h_0\|_{\mathcal{H}}\right) = o_p(1)
\, .
$$

Since $\widehat{g}(z) = (1,z)\widehat{\beta} + \widehat{h}(z)$,
to show uniform consistency of $\widehat{g}$ and $\widehat{g}'$, it
now suffices to show that $\sup_{z\in [0,1]}|\widehat{h}(z)-h_0(z)|$
and $\sup_{z\in[0,1]}|\widehat h '(z)-h_0'(z)|$ are bounded by
$\|\widehat h - h_0\|_{\mathcal{H}}$.  As for
any $h\in\mathcal{H}$, $h'(z)=\int_{0}^z h''(t)dt$,
\begin{equation*}
\label{eq: uniform bound on the 1st derivative of h}
\sup_{z\in[0,1]}|\widehat h'(z)-h_0'(z)|\leq
 \sup_{z\in[0,1]}\int_{0}^z|\widehat{h}''(t)-h_0''(t)|dt\leq
 \int_{0}^1|\widehat{h}''(t)-h_0''(t)|dt\leq \|\widehat h -
 h_0\|_{\mathcal{H}}\, ,
\end{equation*}
from Cauchy-Schwartz inequality.
Since $h(z)=\int_{0}^z h'(t) d t$,  a similar reasoning yields
\begin{equation*}
\label{eq: uniform bound on h}
\sup_{z\in[0,1]}|\widehat h (z)- h_0(z)|\leq
\sup_{z\in[0,1]}\int_{0}^z|\widehat{h}'(t)-h_0'(t)|dt\leq
\sup_{z\in[0,1]}|\widehat h'(z)-h_0'(z)|\, .
\end{equation*}

\textbf{Step 5}.
We now obtain uniform convergence rates.  Assumption \ref{ass:source
condition} allows applying Lemma S3.3(e) which yields
$\|h_\lambda-h_0\|_{\mathcal{H}}=O\left(\lambda^\frac{\gamma \wedge
2}{2}\right)$. Combining with the results of Step 3 gives
$
\|\widehat h - h_0\|_{\mathcal{H}}=O_p\left(\frac{1}{\sqrt{n
\lambda}}+\lambda^{\frac{\gamma \wedge 2}{2}}\right)\text{ and
}\|\widehat \beta - \beta_0\|=O_p\left(\frac{1}{\sqrt{n
\lambda}}+\lambda^{\frac{\gamma \wedge 2}{2}}\right)\, .
$
Use the same arguments as in Step 4 to obtain
uniform convergence rates.
\hfill$\square$\\

\newpage

\renewcommand{\theequation}{S.\arabic{equation}}
\renewcommand{\thesection}{S\arabic{section}}
\renewcommand{\thepage}{S\arabic{page}}

\setcounter{equation}{0}
\setcounter{page}{1}
\setcounter{section}{0}

\begin{center}
{\LARGE \bf Online supplement}
\end{center}

\section{Monte Carlo}
\setcounter{equation}{0}
\setcounter{theorem}{0}
\textcolor{black}{\textbf{Implementation of the Tikhonov estimator.} The Tikhonov optimization problem at the  population
level is
\[
\arg\min_g \| \E \left[ Y|W\right]- \E \left[g(Z) | W \right]\|_{L^2_W}^2  +
\lambda \| g\|_{H^2_Z}^2,
\, .
\]
where $$\| g\|_{H^2_Z}^2=\int_0^1g(z)^2dz + \int_0^1g'(z)^2dz+
\int_0^1g"(z)^2dz$$ is the Sobolev norm of degree $2$.  For the
practical implementation, we follow the approach of
\citet[Section 7]{gagliardini2012tikhonov}. The conditional expectation operator
given $W$ is estimated by kernel smoothing, where  we used a Gaussian
kernel of order 2 and a bandwidth selected using Silverman's rule of thumb, i.e.,
equal to $n^{-1/5}$ times the empirical standard deviation of $W$.
The Sobolev space of order 2 is approximated through the basis of the
first 10 Chebyshev polynomials of the first kind shifted to
$[0,1]$. To select the regularization parameter $\lambda$, we used
two-fold cross-validation, where we minimize the empirical analog of the
loss $\| \E \left[ Y|W\right]- \E \left[g(Z) | W \right]\|_{L^2_W}^2$,
and we use the same grid of values of $\lambda$ as for our smoothing
splines estimator.}

\smallskip
\noindent \textbf{Implementation of the series estimator.} We also considered a series estimator based on a basis of Legendre
polynomials. The main idea is to consider the equality
\[
\E \left[ Y|W\right] f_W(W)  = \int_{}^{}{}  g(Z)  f_{(Z,W)}(Z,W) \, dZ
\, .
\]
The right-hand side of the equation, the function $g$, and the
joint density $f_{(Z, W)}$ are each approximated by a series expansion, respectively
, on $J$, $K$, and $J\times K$ terms. We used a method proposed by
\citet{horowitz2014adaptive}, who considered the case $J=K$ and derived
an adaptive procedure to select  $J$.

\smallskip
\noindent\textcolor{black}{ \textbf{Pre-transformation of variables.}
Since both the Tikhonov and Series methods are designed for variables
belonging to $[0,1]$, we transformed observations of $Z$ and $W$ by
their respective empirical cumulative distribution functions (cdf).
(In unreported simulations, we found that using the true cdfs instead
did not affect our results much.) This implies in particular that even
if the relation between $Z$ and $W$ is linear, the first-stage equation is not linear anymore in the transformed variables.}

\begin{figure}[!ht]
\centering
\begin{subfigure}{0.5\textwidth}
  \centering
  \includegraphics[width=\linewidth]{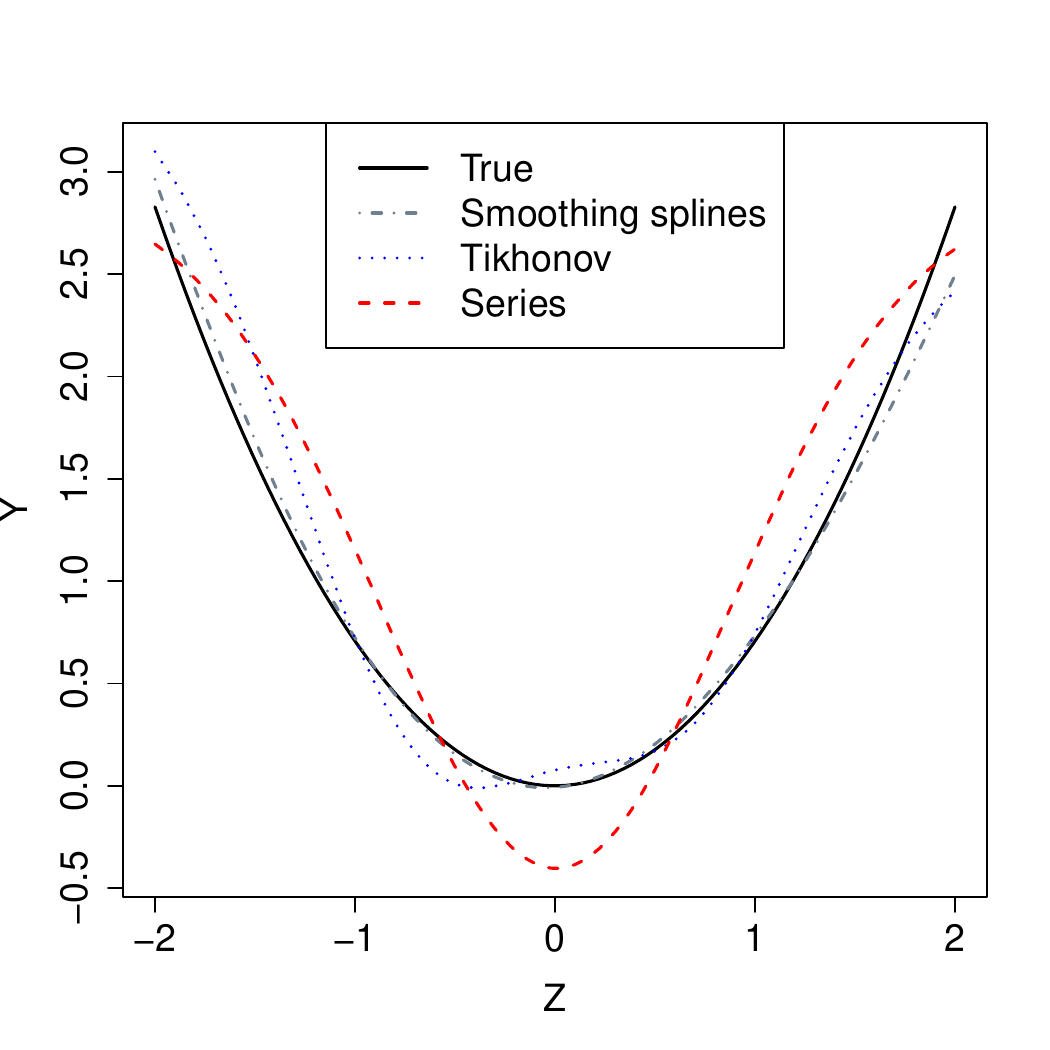}
  \caption{$g_{0,1}$}
  \label{fig6:a}
\end{subfigure}%
\begin{subfigure}{0.5\textwidth}
  \centering
  \includegraphics[width=\linewidth]{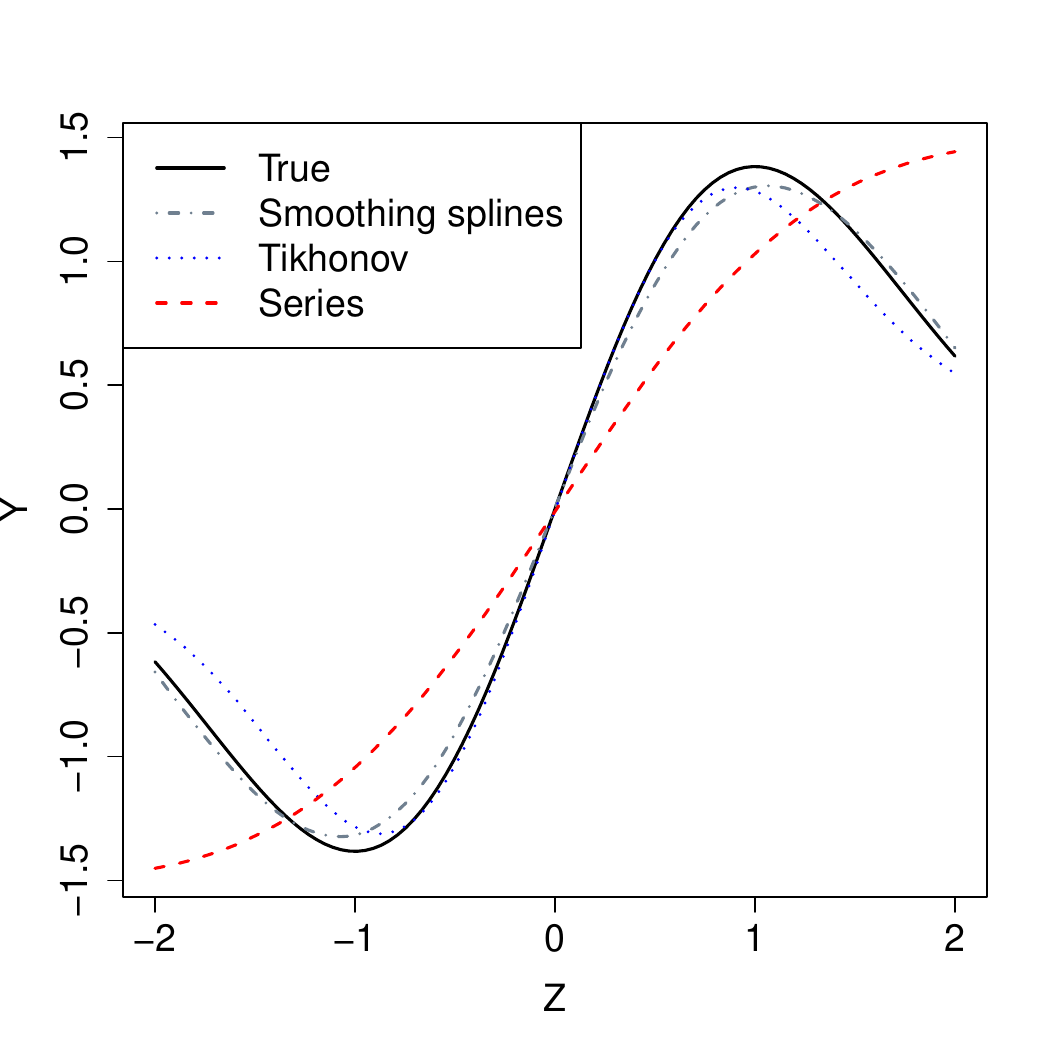}
  \caption{$g_{0,2}$}
  \label{fig6:b}
\end{subfigure}
\begin{subfigure}{0.5\textwidth}
  \centering
  \includegraphics[width=\linewidth]{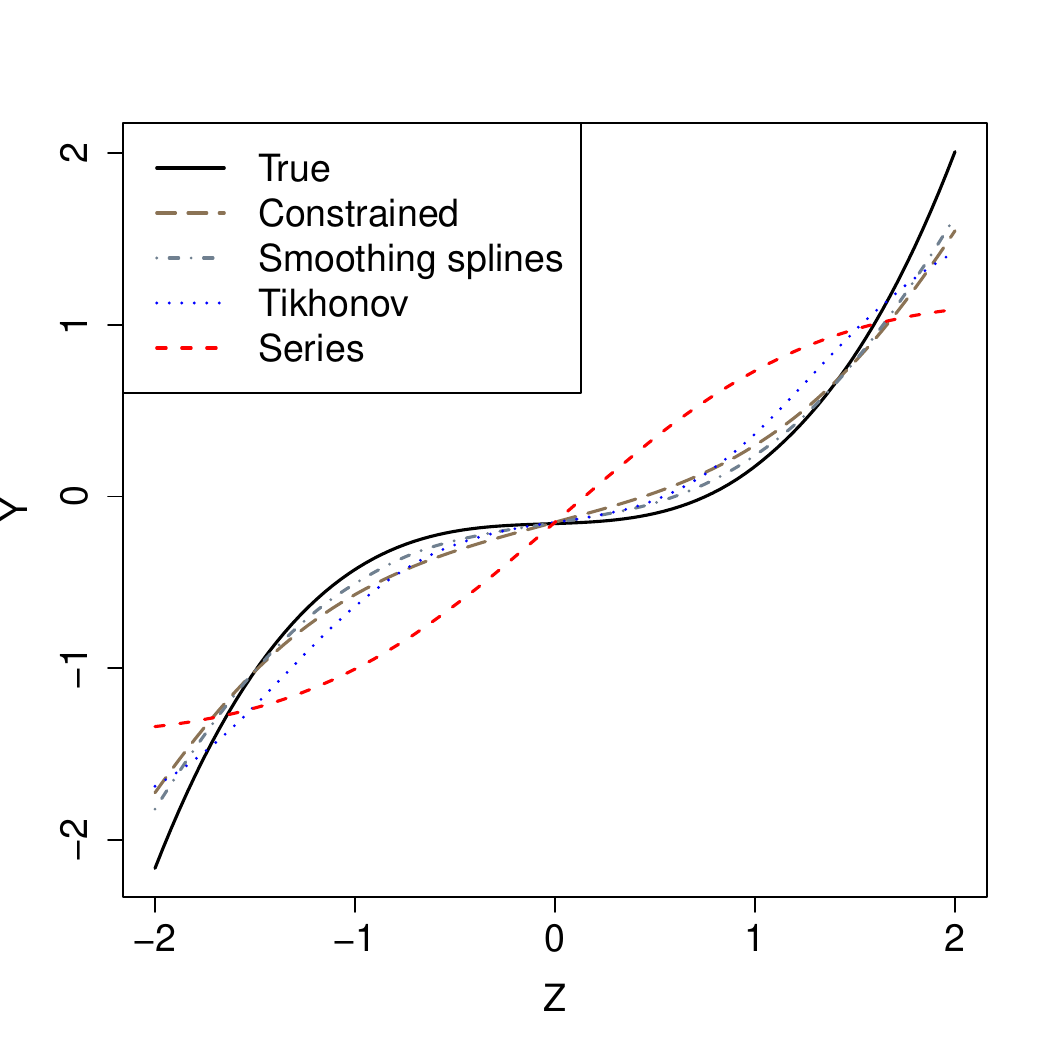}
\caption{$g_{0,3}$}
  \label{fig6:c}
\end{subfigure}%
\caption{True regression function and average estimators for
$n=200$,  $\rho_{\varepsilon V}=0.8$, and $\rho_{ZW}=0.7$. }
\label{fig6}
\end{figure}

\smallskip
\noindent \textbf{Supplementary results.}
Figures \ref{fig6:a}, \ref{fig6:b} and \ref{fig6:c} graph the pointwise average of each
estimator for $n=200$ and $(\rho_{WZ}, \rho_{\varepsilon V} ) =
(0.7,0.8)$. Here, the series estimator is much steeper than the true
quadratic curve $g_{0,1}$, while it fails to fit the sign changes in
the first derivative for $g_{0,2}$.
{\color{black} In all cases, the Tikhonov estimator is pretty close to
the true curves, while the smoothing splines estimator is almost
unbiased.}  From our figures, the degree of smoothing appears to be
quite different among the averaged estimators. However, there is no
clearly accepted way to measure degrees of freedom in nonparametric
instrumental variable regression. The issue is particularly intricate
for our competitors. {\color{black} The Tikhonov estimator depends on a
bandwidth as well as a regularization parameter, and the influence of
each choice on the final estimator is far from clear.} The same comment
applies to the series estimator, which relies on three estimated
nonparametric components. The method proposed by
\citet{horowitz2014adaptive} simplifies the matter by making each
dependent upon a single parameter $J$, but the effect of this choice
on the final estimator remains to be investigated. By contrast, our
smoothing spline estimator depends upon a single regularization
parameter.

\section{Additional proofs}
 \setcounter{equation}{0}
\setcounter{theorem}{0}
\medskip
\textbf{Proof of Proposition 2.1:}

(a) {\em Unicity}.
We begin by studying   $\bm{\Omega}$. Let
$\bm{b}=(b_1,\ldots,b_n)^T\in \mathbb{R}^n$, then
\[
    \bm{b^T \Omega b}=\int_{ }\left|\frac{1}{n}\sum_{i=1}^n b_i
    \exp(\textbf{i}W_i^T t)\right|^2 \mu(dt)\geq 0\, .
\]
Hence  $\bm{b^T\Omega b}=0$ iff
$ (1/n)\sum_{i=1}^n b_i \exp(\mathbf{i}W_i^T t)=0$  for all
$t\in\mathbb{R}^q$.
Define the random vector $(\widetilde{b},\widetilde{\bm{W}})$  that
equals $(b_i,W_i)$ with probability $1/n$, and  $\widetilde{\E}$ the
corresponding expectation.
Then, $\widetilde{\E}[\widetilde{b}\exp(\mathbf{i}\widetilde{\bm{W}}^T t)]=0$
for all $t\in\mathbb{R}^q$. From \cite{Bierens1982}, this implies that
$ \widetilde{\E}[\widetilde{{b}}\,|\,\widetilde{\bm{W}} = \bm{W}_i ]=0$.
Since $ \widetilde{\E}[\widetilde{{b}}\,|\,\widetilde{\bm{W}} = \bm{W}_i]
= b_i$  if all $W_i$s are different,  $b_i=0$ for all
$i=1,\ldots,n$. Hence, $\bm{\Omega}$ is positive definite.

 From \citet[Chapter 2]{green1993nonparametric},
a natural cubic spline is uniquely defined by the vector of its
values at the knots $\bm{g}$, and
 we can write
\[
\int  g'' (z) ^{2} \, dz  = \bm{g^T K g}
\, ,
\]
for a positive semi-definite matrix $\bm{K}$.
Hence, our minimization problem writes
\begin{equation*}
    \min_{\bm{g}} \bm{\left( Y - g\right)^T \Omega  \left( Y-g\right)} +
\lambda \bm{g^T K g}
=
\min_{\bm{g}} \bm{ g^T \left(  \Omega  + \lambda K \right)g} +
2 \ \bm{g^T \Omega Y} + \bm{Y^T\Omega Y}
\, .
\end{equation*}
Since $\bm{ \Omega  + \lambda K }$ is positive-definite for any
$\lambda >0$, the problem is convex and has a unique minimum.

(b) {\em Solution}.
A natural cubic spline $g$ {can also be uniquely written} as
\begin{equation}\label{eq: representation of a cubic spline}
g(z)  =  a_{0} + a_{1} z + \frac{1}{12} \sum_{i=1}^{n}{ \delta_{i}
|z-Z_{i}|^{3}}\, ,
\quad
\sum_{i=1}^{n}{ \delta_{i}} = \sum_{i=1}^{n}{\delta_{i} Z_{i}} = 0
\, ,
\end{equation}
whenever the $Z_i$'s are all different. One can thus write
 $\bm{g  = E \delta + Za}$, with $\bm{Z^T\delta = 0}$.

To show that the above formulation is unique, let us check that $\bm{E
 \delta + Za = 0}$  with $\bm{Z^T\delta = 0}$ implies the nullity of all coefficients.
Our premises yield $\bm{\delta^T (E\delta + Za) = \delta^T E \delta} = 0$.
From  \citet[Section 7.3]{green1993nonparametric},
\begin{equation}
\int  g'' (z) ^{2} \, dz  = \bm{\delta^T {E} \delta} \geq 0
\, .
\label{roughness}
\end{equation}
Hence, since $g''$is continuous, it should be that $g''$ is identically
 zero. Since the increments of the third derivative at the knots are
 $g'''(Z_i^+) - g'''(Z_i^-) = \delta_i$, $i = 1, \ldots n$, this
 implies that $\bm{\delta = 0}$. Finally, $\bm{Za = 0}$ implies $\bm{a
 = 0}$, as $\bm{Z}$ is full rank.

From the uniqueness obtained in Part (a),
\begin{equation}
\left( \bm{Y - Z a - E \delta}  \right)^T \bm{\Omega} \left( \bm{Y} - \bm{Z a} -
\bm{E \delta}  \right)
+ \lambda \bm{\delta^T {E} \delta}
\label{eq:minp}
\end{equation}
admits a unique global minimum
under the constraint  $\bm{Z^T\delta = 0}$. To
characterize such a minimum, consider the Lagrangian
\begin{equation*}
 \left( \bm{Y - Z a - E \delta}  \right)^T \bm{\Omega} \left( \bm{Y} - \bm{Z a} -
\bm{E \delta}  \right)
+ \lambda \bm{\delta^T {E} \delta} + \bm{l}^T \bm{Z}^T\bm{\delta} \,,
\end{equation*}
where $\bm{l}\in \mathbb{R}^2$ is the vector of Lagrange multipliers,
and the associated first-order conditions with respect to $(\bm{\delta}^T,\bm{a}^T)^T$
\begin{equation*}
\left[
\begin{array}{cc}
\bm{E^T \Omega}  & \bm{0}
\\
\bm{Z^T \Omega} & - \lambda \bm{I}
\end{array}
\right]
\left[
\begin{array}{cc}
\widetilde{\bm{E}} & \bm{Z}
\\
\bm{Z^T} & \bm{0}
\end{array}
\right]
\left(
\begin{array}{c}
\bm{\delta}
\\
\bm{a}
\end{array}
\right)
-
\left[
\begin{array}{cc}
\bm{E^T \Omega}  & \bm{0}
\\
\bm{Z^T \Omega} & - \lambda \bm{I}
\end{array}
\right]
\left(
\begin{array}{c}
\bm{Y}
\\
\bm{0}
\end{array}
\right) -
\left(
\begin{array}{c}
\bm{Z}\, \bm{l}
\\
\bm{0}
\end{array}
\right)
=\left(
\begin{array}{c}
\bm{0}
\\
\bm{0}
\end{array}
\right)
\, .
\end{equation*}
One solution to these first-order conditions is given by
$\bm{l}=(0,0)^T$ and $(\bm{\widehat{\delta}}^T,\bm{\widehat{a}}^T)^T$
satisfying
\begin{align}
\left[
\begin{array}{cc}
\widetilde{\bm{E}} & \bm{Z}
\\
\bm{Z^T} & \bm{0}
\end{array}
\right]
\left(
\begin{array}{c}
\bm{\widehat{\delta}}
\\
\bm{\widehat{a}}
\end{array}
\right)
& =
\left(
\begin{array}{c}
\bm{Y}
\\
\bm{0}
\end{array}
\right)
\, ,
\qquad
\widetilde{\bm{E}} = \bm{E} + \lambda \bm{\Omega}^{-1}
\, ,
\label{estimformulap}
\end{align}
and this solution  satisfies the constraint $\bm{Z^T \widehat{\delta}}
= \bm{0}$.
This solution is a strict local minimum if
the second-order sufficient conditions hold
\cite[page 334]{Luenberger2008}, that is  for any non-zero
$\left(\bm{\delta}^T, \bm{a}^T\right)^T $ satisfying $\bm{Z^T
{\delta}} = \bm{0}$,
\begin{equation}
\label{eq:soc}
\left(
\begin{array}{c}
\bm{\delta}
\\
\bm{a}
\end{array}
\right)^{\bm{\top}}
\left[
\begin{array}{cc}
\bm{E^T \Omega E} + \lambda \bm{E} & \bm{E^T \Omega Z}
\\
\bm{Z^T\Omega E} & \bm{Z^T\Omega Z}
\end{array}
\right]
\left(
\begin{array}{c}
\bm{\delta}
\\
\bm{a}
\end{array}
\right)
 =
\bm{ \left( E\delta + Za \right)^T \Omega \left( E\delta + Za \right)}
+ \lambda \bm{\delta^T E \delta}
> 0
\, .
\end{equation}
As $\bm{\Omega}$ is positive definite, and using (\ref{roughness}), (\ref{eq:soc}) is
non-negative, and is zero iff $\bm{ E\delta + Za} = \bm{0}$. But this
would imply $\left(\bm{\delta}^T, \bm{a}^T\right)^T= \bm{0}$ as shown above.

The right-hand side matrix in (\ref{estimformulap}) is full rank as $\bm{Z}$ is full
rank and $\lambda > 0$. Indeed, assume
\[
\left[
\begin{array}{cc}
\bm{E} + \lambda  \bm{\Omega}^{-1} &  \bm{Z}
\\
\bm{Z^T} & \bm{0}
\end{array}
\right]
\left(
\begin{array}{c}
\bm{\delta}
\\
\bm{a}
\end{array}
\right)
=
\left(
\begin{array}{c}
\bm{0}
\\
\bm{0}
\end{array}
\right)
\, ,
\]
this implies $\bm{Z^T\delta} = \bm{0}$ and
$\bm{0}= \bm{\delta^T}
\left[ \left(\bm{E} + \lambda \bm{\Omega}^{-1} \right) \bm{\delta}  +  \bm{Z a} \right] =
\bm{\delta^T}  \bm{E} \bm{ \delta} +
\lambda \bm{\delta^T}  \bm{\Omega}^{-1}  \bm{ \delta}$.
From (\ref{roughness}) and the positive-definiteness of $\bm{\Omega}$,
this yields $\bm{\delta = 0}$, and in turn $\bm{a = 0}$.

To obtain the values at the knots $\widehat{\bm{g}}$, note that the inverse of the matrix
in (\ref{estimformulap}) is
\begin{equation}
\left[
\begin{array}{cc}
\widetilde{\bm{E}}^{-1} \left(\bm{I}-\bm{P}\right)
  &
 \widetilde{\bm{E}}^{-1} \bm{Z} \left(\bm{Z}^T \widetilde{\bm{E}}^{-1} \bm{Z} \right)^{-1}
\\
\left(\bm{Z}^T \widetilde{\bm{E}}^{-1} \bm{Z} \right)^{-1}  \bm{Z}^T \widetilde{\bm{E}}^{-1}
 &  - \left(\bm{Z}^T \widetilde{\bm{E}}^{-1} \bm{Z} \right)^{-1}
\end{array}
\right]
\, ,
\label{invmat}
\end{equation}
where $\bm{P} $ is the oblique projection on
the span of $\bm{Z}$  along the direction spanned by vectors $\bm{h}$
such that  $\bm{Z}^T  \widetilde{\bm{E}}^{-1} \bm{h} = 0$.  Hence,
\begin{align*}
\left(
\begin{array}{c}
\bm{E} \widehat{\bm{\delta}}
\\
\bm{Z} \widehat{\bm{a}}
\end{array}
\right)
&  =
\left[
\begin{array}{c}
\bm{E} \widetilde{\bm{E}}^{-1} \left(\bm{I}-\bm{P}\right) \bm{Y}
\\
\bm{P} \bm{Y}
\end{array}
\right]
\, .
\end{align*}
Use  $\widehat{\bm{g}}  =  \bm{Z \widehat{a}} + \bm{E
\widehat{\delta}}$  to obtain the desired result.
\hfill$\square$

      \medskip

\noindent\textbf{Proof of Theorem 6.1:}
We first prove that Assumption \ref{ass:completeness pl} implies
\begin{enumerate}
    \item[(a)] $\operatorname{D}$ and $\A$ are injective,
    \item[(b)] $\mathcal{R}(\A)\cap \mathcal{R}(\operatorname{D})=\{0\}$\,,
    \item[(c)] $\bm{L}$ is full column rank with probability approaching one,
\end{enumerate}
where $\mathcal{R}(\A)$ denotes the range of $\A$, with $\A$ defined
in (\ref{eq: definitions of of A and B}), and
$\mathcal{R}(\operatorname{D})$ denotes the range of
$\operatorname{D}$, with $\operatorname{D}$ is defined in (\ref{eq:
definition of D}).

The proof of condition (a) uses arguments contained in the proof of
Step 1 of Theorem \ref{th.cvrates} in the main text. For completeness, we also provide it here. Given
$\beta\in\mathbb{R}^2$ and $h\in\mathcal{H}$ (with $\mathcal{H}$
defined in Section \ref{sec: Asymptotic analysis}), for
$g(z)=(1,z)\beta+h(z)$ we have that $g\in\mathcal{G}$. So,
\begin{align*}
    (\beta,\gamma)=0\,,\,h=0\quad \Leftrightarrow & \quad \E[X^T\gamma+(1,Z)\beta+h(Z)|W]=0\\
    \Leftrightarrow & \quad \E[(X^T\gamma+(1,Z)\beta+h(Z))\exp(\mathbf{i}W^T t)]=0\, , \quad \forall t\in\mathbb{R}^p\\
    \Leftrightarrow & \quad  \operatorname{D}(\gamma,\beta)+\A h=0\, ,
\end{align*}
where the first equivalence follows from Assumption
\ref{ass:completeness pl}, the second equivalence from
\cite{Bierens1982}, and the last equivalence from the definition of
$\operatorname{D}$ and $\A$.  Since $\A 0=0$, by the above display
$\operatorname{D}(\gamma,\beta)=0$ implies $(\gamma,\beta)=0$. So,
$\operatorname{D}$ is injective. Similarly, the above display also
ensures that $\A$ is injective. Thus, condition (a) is proved.  To
prove condition (b), consider an element belonging to
$\mathcal{R}(\A)\cap \mathcal{R}(\operatorname{D})$, say $\A
h=\operatorname{D}(\gamma,\beta)$. Then,
$\operatorname{D}(\gamma,\beta)+A(-h)=0$ and by the above display, we
get that $(\gamma,\beta)=0$ and $h=0$. So, condition (b) is proved.
Let us finally show condition (c). Recall that $\bm{L}$ is the
$n\times (q+2)$ matrix whose $i$th row is $(1,Z_i,X_i^T)$. Take
$(\gamma,\beta)$ such that $(1,Z)\beta+X^T\gamma=0$. Then,
$\operatorname{D}(\gamma,\beta)=0$, and by the injectivity of
$\operatorname{D}$, we obtain that $\gamma=0$ and $\beta=0$. Thus,
$(1,Z,X^T)$ are linearly independent. This implies that
$\E[(1,Z,X^T)^T\,(1,Z,X^T)]$ is full rank. Since $(1/n)\bm{L}^T
\bm{L}$ $=\sum_{i=1}^n (1,Z_i,X^T_i)^T (1,Z_i,X^T_i)$
$=\E[(1,Z,X^T)^T\,(1,Z,X^T)]+o_p(1)$, we get that $\bm{L}^T \bm{L}$
is full rank with probability approaching one. So $\bm{L}$ is full
column rank with probability approaching one and condition (c) is
proved. \\ Given conditions (a), (b), and (c), the proof of Theorem
\ref{th.cvrates pl} proceeds along the same arguments as the proof of
Theorem \ref{th.cvrates} in the main text.
 \hfill$\square$\\

\medskip
\section{Auxiliary lemmas}

The following lemma is  from \citet[Lemma A.1]{florens2011identification}.
\begin{lemma}
\label{lm.bounds}
Consider two Hilbert spaces  $\mathcal{X}$ and $\mathcal{Y}$  and
 a linear compact operator $\operatorname{K}:\mathcal{X}\mapsto
 \mathcal{Y} $. Then there are universal constants $c$ and $c'$ such that
(a) \labeltext{a}{boundii} $||\lambda(\lambda \operatorname{I} +
\operatorname{K}^* \operatorname{K})^{-1}||_{op}\leq c$;
(b) \labeltext{b}{boundiii} $||(\lambda \operatorname{I} + \operatorname{K}^*
\operatorname{K} )^{-1}\operatorname{K}^* ||_{op}\leq
\frac{c'}{\sqrt{\lambda}}$.
\end{lemma}

\begin{lemma}
\label{lm.cvop}
Under Assumptions  \ref{ass:square
integrability} and \ref{ass:completeness},
(a) \labeltext{a}{cratei} the operators $\A$ and $\T$ are compact;
(b) \labeltext{b}{crateii} $\|\widehat{\B}-\B\|_{op}=O_p(n^{-1/2})$;
(c) \labeltext{c}{crateiii} $\|(\widehat{\B}^*\widehat{\B})^{-1}- (\B^*
    \B)^{-1}\|_{op} = O_p(n^{-1/2})$ and $\|\widehat{\M}-\M\|_{op}=O_p(n^{-1/2})$;
 (d) \labeltext{d}{crateiv} $\|\widehat{\A}-\A\|_{op}=O_p(n^{-1/2})$;
 (e) \labeltext{e}{cratev} the operator $\widehat{\T}$ is compact;
(f) \labeltext{f}{cratevi} $\|\widehat{\T}-\T\|_{op}=O_p(n^{-1/2})$;
(g) \labeltext{g}{cratevii} $\left\|\widehat{r}-r\right\|_\mu=O_p(n^{-1/2})$;
(h) \labeltext{h}{crateviii}
    $\|\widehat{\M}\widehat{r}-\widehat{\T}h_0\|_\mu=O_p(n^{-1/2})$.
\end{lemma}

\noindent\textbf{Proof of Lemma \ref{lm.cvop}:}
\eqref{cratei}. Let us show $\A$ is compact by  compact embedding.
Define  $\widetilde{\A}$ as the extension of $A$ to $L^2([0,1])$,
where $L^2([0,1])$ is the space of real-valued squared-integrable
functions on $[0,1]$,  i.e. $\widetilde{\A} h =  \E[h(Z)e^{\mathbf{i}W^\top \cdot}]$ for
any $h \in L^2([0,1])$.
For all $h\in\mathcal{H}$, we have $h(z)=\int_0^z\int_0^x h''(t)dt
dx$, so that
\begin{align*}
||h||_{L^2[0,1]}^2 =\int_0^1|h(z)|^2 dz & \leq
\sup_{z\in[0,1]}| h (z)| \leq
\sup_{z\in[0,1]} \left|\int_0^z h'(t)dt \right|
\leq
\sup_{t\in[0,1]}  |h'(t)|
\\ &
\leq
\sup_{t\in[0,1]} \left|\int_0^t h''(u)du \right|
\leq
\int_0^1 |h''(u)|du
\leq \| h|\|_{\mathcal{H}}^2
\, .
\end{align*}
Therefore, every bounded set on
$(\mathcal{H},||\cdot||_{\mathcal{H}})$ is also a bounded set on
$(L^2([0,1]),||\cdot||)$. Hence, compactness of  $\widetilde{\A}$
implies compactness of $\A$. Now for any $h\in L^2[0,1]$,
\begin{equation*}
  (\widetilde{\A}h)(t)=\E[h(Z)\,\E[\exp(\textbf{i}W^T t)|Z]]=\int
  h(z)\, \E[\exp(\textbf{i}W^T t)|Z=z]\, f_Z(z)\, dz\, ,
\end{equation*}
where
\begin{equation*}
    \int \left|\E[\exp(\textbf{i}W^T t)|Z=z]\right|^2 \mu(t) f_Z(z)\, dt\, dz\leq 1\, ,
\end{equation*}
as $|\exp(\textbf{i}\cdot)|\leq 1$.
Since $\widetilde{\A}$ is an  integral operator whose kernel is
Hilbert-Schmidt, i.e. squared integrable, we can   apply
\citet[Proposition 2.1]{Busby1972}
to conclude that $\widetilde{\A}$ is compact.

Let us now show that $\T$ is compact. The range of $\B$,
$\mathcal{R}(\B)$, is finite dimensional, linear, and closed.  $\P$ is
the orthogonal projection onto $\mathcal{R}(\B)$, and is thus bounded
by \citet[Theorem 13.3]{kress1999linear}.  Hence, $\M=I-\P$ is bounded
as well.  Since $\T= \M \A$ is the composition of a bounded and a
compact operator, it is  compact by \citet[Theorems 2.14 and
2.16]{kress1999linear}.

\eqref{crateii}. For $\beta\in \R^2$, we have
\begin{align*}
\|(\widehat{\B}-\B)\beta\|_{\mu}^2&=\int
|(\E_n-\E)[\exp(\mathbf{i}W^\top t)(1,Z)]\beta|^2\mu(dt)
\, ,
\end{align*}
where $\E_n$ denotes the empirical expectation. By the  Cauchy-Schwarz inequality,
\begin{align*}
\E\|\widehat{\B}-\B\|_{op}^2
\le & \E\left[\int (|(\E_n-\E)[\exp(\mathbf{i}W^\top
t)]|^2+|(\E_n-\E)[Z\exp(\mathbf{i}W^\top t)]|^2)\mu(dt)\right].
\end{align*}

Since  data are i.i.d.,
\begin{align*}
E\left[ |(\E_n-\E)[Z\exp(\mathbf{i}W^\top t)]|^2\right]
& = \E\left[\left|n^{-1}\sum_{i=1}^n Z_i\exp(\mathbf{i}W_i^\top t) -
\E[Z\exp(\mathbf{i}W^\top t)]\right|^2\right]
\\
& =
\operatorname{Var}\left(\frac{1}{n}\sum_{i=1}^n Z_i
\exp(\textbf{i}W^T_i t)\right)
\\
& = n^{-1}(\E[|Z\exp(\mathbf{i}W^\top
t)|^2]-|\E[Z\exp(\mathbf{i}W^\top t)]|^2)
= O(n^{-1}) \, ,
\end{align*}
as $|Z\exp(\mathbf{i}W^\top t)|\leq 1$ for all
$t\in\mathcal{T}$. Similarly, $\E\, |(\E_n-\E)[\exp(\mathbf{i}W^\top
t) ]|^2 = O(n^{-1})$.
This implies $\E\|\widehat{B}-B\|_{op}^2  = O(n^{-1})$, and
by Markov's inequality,  $\|\widehat{\B}-\B\|_{op}^2=O_p(n^{-1}). $

\eqref{crateiii}.
From \citet[Theorem 2.7-8]{kreyszig1978introductory}, as $\B$ is a
linear operator with a finite dimensional domain, it is  bounded, and
$\|\B\|_{op} < \infty$. Also $\|\B^{*}  \|_{op} =
\| \B \|_{op}$ and $\|\B^{*} \B \|_{op} = \| \B \|_{op}^2$.  The operator $\B^*\B$ maps $\R^2$ into $\R^2$, and is thus a matrix.
From  \eqref{crateii}, $\|\widehat{\B}\|_{op}$ and
 $\|\widehat{\B}^{*}\|_{op}$ are $o_p(1)$, and
\begin{align*}
\|\B^*\B-\widehat{\B}^*\widehat{\B}\|_{op}
&=
\| (\B^*-\widehat{\B}^*) B + \widehat{\B}^{*} (\B - \widehat{\B}) \|_{op}
 \\
& \leq
\|\B^*-\widehat{\B}^*\|_{op}\, \|\B\|_{op} + \|\widehat{\B}^*\|_{op}\,
\|\B-\widehat{\B}\|_{op} = o_p(n^{-1/2})
\, .
\end{align*}
 Since $\B$ is injective, $\B^*\B$ is invertible, $(\B^*\B)^{-1}$
 exists and is bounded. By the continuous mapping theorem,
 $\|(\widehat{\B}^*\widehat{\B})^{-1}-(\B^*\B)^{-1}\|_{op} = o_p(1)$.
Hence $\|(\widehat{\B}^*\widehat{\B})^{-1}\|_{op}\leq
\|(\widehat{\B}^*\widehat{\B})^{-1}-(\B^*\B)^{-1}\|_{op}+\|(\B^*\B)^{-1}\||_{op}=o_p(1)$.
Moreover,
\begin{align*}
    \|(\widehat{\B}^*\widehat{\B})^{-1} - (\B^*\B)^{-1} \|_{op} = &
    \|(\widehat{\B}^*\widehat{\B})^{-1}
    (\B^*\B-\widehat{\B}^*\widehat{\B}) (\B^*\B)^{-1} \|_{op}
    \\
    \leq &
    \| (\widehat{\B}^*\widehat{\B})^{-1}\|_{op} \,
    \|\B^*\B-\widehat{\B}^*\widehat{\B}\|_{op}\,
    \|(\B^*\B)^{-1}\|_{op} = o_p(n^{-1/2})
    \,.
\end{align*}

For  the difference between  $\M=\I-\B(\B^*\B)^{-1}\B^*$ and
$\widehat{\M}=\operatorname{I}-\widehat{\B}(\widehat{\B}^*\widehat{\B})^{-1}\widehat{\B}^*$,
we have
\begin{align*}
    \|\widehat{\M} - \M\|_{op}=&\|
    (\widehat{\B}-\B)(\widehat{\B^*}\widehat{\B})^{-1}\widehat{\B}^* +
    \B [(\widehat{\B}^* \widehat{\B})^{-1}- (\B^*\B)^{-1}]
    \widehat{\B}^* + \B (\B ^* \B)^{-1}[\widehat{\B}^* -\B^*]\|_{op}
    \\
    \leq & \|\widehat{\B}-\B\|_{op}\,
    \|(\widehat{\B}^*\widehat{\B})^{-1}\|_{op}
    \, \|\widehat{\B}^*\|_{op} + \|\B\|_{op} \,
    \|(\widehat{\B}^*\widehat{\B})^{-1}-(\B^* \B)^{-1}\| \,
    \|\widehat{\B}^*\|_{op} \\
   & + \|\B\|_{op} \, \|(\B^*\B)^{-1}\|_{op} \,
    \|\widehat{\B}^*-\B^*\|_{op} = o_p(n^{-1/2})
    \, .
\end{align*}

\eqref{crateiv}.
Recall that for any $h\in\mathcal{H}$, $h(z)=\int_{0}^z \int_{0}^x
h''(u) du dx$. Thus,
\begin{align}
\label{eq: expression of Ahat as a kernel operator}
    (\widehat{\A}h)(t)=& \frac{1}{n}\sum_{i=1}^n h(Z_i)
    \exp(\textbf{i}W_i^T t)=\frac{1}{n}\sum_{i=1}^n
    \int_{0}^{Z_i}\int_{0}^x h''(u) \,du\, dx\, \exp(\textbf{i}W_i^T
    t)\nonumber \\
    =& \int_{[0,1]^2}h''(u)\left[\frac{1}{n}\sum_{i=1}^n
    \1(0<u<x)\,\1(0<x<Z_i)\,\exp(\textbf{i}W_i^T t)\right]\,du\,dx
    \nonumber \\
    =& \int_{0}^1 h''(u)\left[\int_{0}^1 \frac{1}{n}\sum_{i=1}^n
    \1(u<x<Z_i)\,\exp(\textbf{i}W_i^T t)\, dx\right]\, du \nonumber \\
    =& \int_{0}^1 h''(u)\, \widehat k(u,t)\, du\, ,
\end{align}
where $\widehat{k}(u,t)$ is defined implicitly above.
Exchanging the empirical measure with the population probability
and using the same steps as above yield
\begin{align}\label{eq: expression of A as a kernel operator}
    (\A h)(t)=\int_{0}^1 h''(u)\left[\int_{0}^1
    \E\{\1(u<x<Z)\,\exp(\textbf{i}W^T t)\}\, dx\right]=\int_{0}^1
    h''(u)\, k(u,t)\, du\, .
\end{align}
where $k(u,t) = \E \widehat{k}(u,t)$ is defined implicitly above.
Next,
\begin{align*}
    \|\widehat \A -
    \A\|_{op}^2=&\sup_{h\in\mathcal{H}\,\|h\|_{\mathcal{H}}=1}\|\widehat
    \A h - \A h \|^2_\mu =\sup_{h\in\mathcal{H}\,\|h\|_{\mathcal{H}}=1}
    \int_{ }\left|(\widehat{\A}h)(t)-(\A h)(t)\right|^2\mu(dt)\\
    =& \sup_{h\in\mathcal{H}\,\|h\|_{\mathcal{H}}=1} \int_{
    }\left|h''(u)\,\left[\widehat{k}(u,t)-k(u,t)\right]\,du\,\right|^2
    \mu(dt)\\
    \leq & \sup_{h\in\mathcal{H}\,\|h\|_{\mathcal{H}}=1} \int_{
    }\left(\int_{0}^1
    |h''(u)|^2du\,\int_{0}^1\left|\widehat{k}(u,t)-k(u,t)\right|^2\,du\right)
    \mu(dt)\\
    =
    &\sup_{h\in\mathcal{H}\,\|h\|_{\mathcal{H}}=1}\|h\|_{\mathcal{H}}
    \int_{
    }\left(\int_{0}^1\left|\widehat{k}(u,t)-k(u,t)\right|^2\,du\right)
    \mu(dt)\\
    =& \int_{ [0,1]\times \mathbb{R}^q
    }\left|\widehat{k}(u,t)-k(u,t)\right|^2\,du\otimes \mu(dt)
    \\
\Rightarrow
    \E \|\widehat \A - \A\|^2_{op}
    \leq  & \int_{ [0,1]\times \mathbb{R}^q } \E
    \left|\widehat{k}(u,t)-k(u,t)\right|^2\,du\otimes \mu(dt)\, .
\end{align*}
Now,
\begin{align*}
     \E\left|\widehat{k}(u,t)-k(u,t)\right|^2
     = &\E\left|  \int_{0}^1(\E_n-\E)\1(u<x<Z)\exp(\textbf{i}W^T t)\,
     dx\right|^2
\\  \leq &
 \int_{0}^1 \E \left|(\E_n-\E)\1(u<x<Z)\exp(\textbf{i}W^T t)\right|^2
 \, dx
 \nonumber \\
 =&
 \int_{0}^1 \operatorname{Var}\left(\frac{1}{n}\sum_{i=1}^n
 \1(u<x<Z_i)\exp(\textbf{i}W_i^T t)\right)\, dx
 \nonumber\\
 =& \frac{1}{n} \int_{0}^1  \operatorname{Var}
 \left(\1(u<x<Z)\exp(\textbf{i}W^T t)\right)\, dx  = O(n^{-1})
\, .
\end{align*}
Use Markov's inequality to obtain the desired result.

\eqref{cratev}.
By reasoning as in the proof of \eqref{cratei}, compactness of
$\widehat \T =\widehat{\M} \widehat{\A}$  follows if
 $\widehat{\M}$ is bounded and $\widehat \A$ is compact.
The first claim is shown following similar arguments as in
\eqref{cratei}.
 To obtain compactness of $\widehat \A$, we will use Theorem 8.1-4 in
 \cite{kreyszig1978introductory} stating that a bounded operator with a finite dimensional range is
compact.
As
\begin{align*}
 \widehat{\A}h=\sum_{i=1}^n h(Z_i)\,\frac{1}{n}\exp(W_i^T \cdot)\, \in
 \text{Span}\left(\frac{1}{n}\exp(W_1^T
 \cdot),\ldots,\frac{1}{n}\exp(W_n^T \cdot)\right)
 \, ,
\end{align*}
the range of $\widehat \A$ is finite dimensional for all $n$.
Moreover, using (\ref{eq: expression of Ahat as a kernel operator})
\begin{align*}
\|\widehat{\A}h\|^2_\mu = &
\int \left| \int_{0}^1 h''(u)\, \widehat k(u,t)\, du \right|^2 \mu(dt)
\leq   \|h\|_\mathcal{H}^2 \sup_{u,t} | \widehat k(u,t)|^{2}
\, du
 \leq \|h\|_\mathcal{H}^2
\, ,
\end{align*}
as $| \widehat k(u,t)| \leq 1$. Hence, $\|\widehat{\A}\|_{op} \leq 1$,
and $\widehat \A$ is compact.

\eqref{cratevi}.
Since
$\widehat{\T}-\T = (\widehat{\M}-\M) \widehat \A + \M (\widehat{\A} - \A)$,
 the result follows from \eqref{crateiii}, \eqref{crateiv}, and the
 fact that $\M$ is bounded.

\eqref{cratevii}. The proof  is analogous to the proof of
 \eqref{crateii}.

\eqref{crateviii}.
Write $\widehat{\M}\widehat r - \M
r=(\widehat{\M}-\M)\widehat{r}+\M(\widehat{r}-r)$, and use
$\|\widehat{\M}-\M\|_{op}=O_p(n^{-1/2})$,
 $\|\widehat{r}-r\|_\mu=O_p(n^{-1/2})$, and
$\|\M\|_{op}<\infty$ from previous items to obtain
 $\|\widehat{\M} \widehat r - \M r\|_\mu=O_p(n^{-1/2})$.
Use \eqref{cratevi} above to get $\|(\widehat{\M} \widehat r- \widehat
\T h_0) - (\M r-\T h_0)\|_\mu=O_p(n^{-1/2})$, and note that $\M r-\T
h_0=0$.
   \hfill$\square$

\vspace{0.5cm}
\textcolor{black}{
To present the following lemma, we recall the concept of ``uniform
boundedness in probability". Let
$\Lambda=[\underline{\lambda}_n,\overline{\lambda}_n]$ be such that $n
\underline{\lambda}_n\rightarrow \infty$ and
$\overline{\lambda}_n\rightarrow 0$. For a positive deterministic
function $b$ depending on $n$ and $\lambda\in\Lambda$, say
$b(n,\lambda)$, and a sequence of random variables $(U_n(\lambda))_n$
possibly depending on $\lambda$, we say that
$U_n(\lambda)=O_P(b(n,\lambda))$ uniformly in $\lambda\in\Lambda$ if
$\sup_{\lambda\in\Lambda}b(n,\lambda)|U_n(\lambda)|=O_P(1)$.
\begin{lemma}\label{lm.uniforminlambda}
Let Assumptions \ref{ass:square integrability} and \ref{ass:completeness} hold. Let
$\Lambda=[\underline{\lambda}_n,\overline{\lambda}_n]$ be such that $n
\underline{\lambda}_n\rightarrow \infty$ and
$\overline{\lambda}_n\rightarrow 0$.
Then, (a) $(\T^* \T + \lambda \I )^{-1}\T^* (\widehat{\M}\widehat r-
    \widehat{\T} h_0)\, $,  (b) $
    (\T^* \T + \lambda \I )^{-1}(\widehat{\T}^*-
    \T^*)(\widehat{\M}\widehat r - \widehat{\T}h_0)\,$, (c) $
    [(\widehat{\T}^* \widehat{\T}+\lambda \I)^{-1} $ $- (\T^* \T
    + \lambda \I )^{-1} ]\widehat{\T}^* (\widehat{\M}\widehat r
    - \widehat{\T}h_0)\, $, and (d) $
    (\widehat{\T}^* \widehat{\T}+\lambda \I)^{-1}\widehat{\T}^*
    \widehat{\T}h_0 - $ $(\T^* \T+\lambda \I)^{-1}\T^*
    \T h_0$ are all $O_P(1/\sqrt{n \lambda})$ uniformly in $\lambda\in\Lambda$.
    (e) If moreover Assumption \ref{ass:source condition} holds, then $\|(\T^* \T+\lambda \I)^{-1}\T^*
    \T h_0 - h_0\|_{\mathcal{H}}=O(\lambda^{(\gamma\wedge 2)/2})$ uniformly in $\lambda\in\Lambda$.
\end{lemma}
\textbf{Proof of Lemma \ref{lm.uniforminlambda}:} (a). We have that uniformly in $\lambda\in\Lambda$
\begin{align*}
    \|(\T^* \T + \lambda \I )^{-1}\T^* (\widehat{\M}\widehat r-
    \widehat{\T} h_0)\|_{\mathcal{H}}\leq \|(\T^* \T + \lambda \I
    )^{-1}\T^*\|_{op}\,\|\widehat{\M}\widehat r -
    \widehat{\T}h_0\|_{\mu}=O_p\left(\frac{1}{\sqrt{n
    \lambda}}\right)\, .
\end{align*}
Indeed,  $\T$ is a compact operator  from Lemma S3.2(a),
$\|(\T^* \T + \lambda \I )^{-1}\T^*\|_{op}\leq c'/\sqrt{\lambda}$ from Lemma S3.1(b), and
$\|\widehat{\M}\widehat r - \widehat{\T}h_0\|_{\mu}=O_p(1/\sqrt{n})$
from Lemma S3.2(h).\\
(b). Next,
\begin{align*}
    \|(\T^* \T + \lambda \I )^{-1}(\widehat{\T}^*-
    \T^*)(\widehat{\M}\widehat r - \widehat{\T}h_0)\|_{\mathcal{H}}\leq& \|(\T^* \T + \lambda \I )^{-1}\|_{op}\,\|\widehat{\T}^*-\T^*\|_{op}\,\|\widehat{\M}\widehat
    r - \widehat{\T}h_0\|_{\mu}\\
    &=O_p\left(\frac{1}{n \lambda}\right)\,
\end{align*}
uniformly in $\lambda\in\Lambda$,
as  $\|(\T^* \T + \lambda \I )^{-1}\|_{op}\leq
c/\lambda$ from Lemma S3.1(a),
$\|\widehat{\T}-\T\|_{op}=O_p(1/\sqrt{n})$  and $\|\widehat{\M}\widehat r -
\widehat{\T}h_0\|_{\mu}=O_p(1/\sqrt{n})$ from Lemma
S3.2(f) and (h). \\
(c). Next,
\begin{align*}
    \|[(\widehat{\T}^* \widehat{\T}+\lambda \I)^{-1} -& (\T^* \T
    + \lambda \I )^{-1} ]\widehat{\T}^* (\widehat{\M}\widehat r
    - \widehat{\T}h_0)\|_{\mathcal{H}}\\
    \leq &\|(\widehat{\T}^* \widehat{\T}+\lambda
    I)^{-1}\widehat{\T}^* - (\T^* \T + \lambda \I
    )^{-1}\widehat{\T}^*\|_{op}\, \|\widehat{\M}\widehat r -
    \widehat{\T}h_0\|_{\mu}\nonumber \\
    \leq & \|(\widehat{\T}^* \widehat{\T}+\lambda
    I)^{-1}\widehat{\T}^*-(\T^* \T + \lambda \I )^{-1}\T^*-(\T^* \T +
    \lambda \I )^{-1}(\widehat{\T}^*-\T^*)\|_{op}\,\\
    &\cdot
    \|\widehat{\M}\widehat r - \widehat{\T}h_0\|_{\mu}\nonumber \\
    \leq & \|(\widehat{\T}^* \widehat{\T}+\lambda
    I)^{-1}\widehat{\T}^* -(\T^* \T + \lambda \I )^{-1}\T^* \|_{op}\,
    \|\widehat{\M}\widehat r - \widehat{\T}h_0\|_{\mu}\nonumber \\
    &+ \|(\T^* \T + \lambda \I )^{-1}\|_{op}\,\|\widehat{\T}^* -
    \T^*\|_{op}\,\|\widehat{\M}\widehat r -
    \widehat{\T}h_0\|_{\mu}\nonumber = O_p\left(\frac{1}{\sqrt{n \lambda}}\right)\,
\end{align*}
uniformly in $\lambda\in\Lambda$.
Indeed,  $\widehat{\T}$ and $\T$ are
compact operators  from Lemma S3.2(a) and (e), so
$\|(\widehat{\T}^* \widehat{\T}+\lambda \I)^{-1}\widehat{\T}^* -(\T^*
\T + \lambda \I )^{-1}\T^*\|_{op}\leq 2 c'/\sqrt{\lambda}$  and
$\|(\T^* \T+\lambda \I)^{-1}\|_{op}\leq
c/\lambda$ by Lemma S3.1. Moreover,
$\|\widehat{\T}^*-\T^*\|_{op}=O_p(1/\sqrt{n})$ from Lemma
S3.2(f).\\
(d). Finally,
\begin{align*}
\|(\widehat{\T}^* \widehat{\T}+\lambda \I)^{-1}&\widehat{\T}^*
    \widehat{\T}h_0 -  (\T^* \T+\lambda \I)^{-1}\T^*
    \T h_0\|_{\mathcal{H}}\\
    =&\|(\widehat{\T}^*\widehat{\T}+\lambda \I)^{-1}
\widehat{\T}^*\widehat{\T} h_0 - (\T^*\T + \lambda \I)^{-1} \T^*\T
h_0\|_{\mathcal{H}}
\nonumber\\
=& \|(\widehat{\T}^*\widehat{\T}+\lambda \I)^{-1}
(\widehat{\T}^*\widehat{\T} +\lambda \I - \lambda \I) h_0
- (\T^*\T + \lambda \I )^{-1}(\T^*\T+\lambda \I - \lambda \I )
h_0\|_{\mathcal{H}}
\nonumber\\
=& \|\lambda [(\T^*\T + \lambda \I )^{-1} -
(\widehat{\T}^*\widehat{\T}+\lambda \I)^{-1}] h_0\|_{\mathcal{H}}
\nonumber\\
=& \|\lambda(\widehat{\T}^*\widehat{\T} + \lambda \I )^{-1}
[\widehat{\T}^* \widehat{\T} - \T^*\T] (\T^*\T+\lambda \I)^{-1}
h_0\|_{\mathcal{H}}
\nonumber\\
= & \|\lambda(\widehat{\T}^*\widehat{\T} + \lambda \I )^{-1}
[\widehat{\T}^* (\widehat{\T}-\T) + (\widehat{\T}^*-\T^*)\T]
(\T^*\T+\lambda \I)^{-1} h_0\|_{\mathcal{H}}
\nonumber\\
\leq & \|(\widehat{\T}^*\widehat{\T}+\lambda \I)^{-1}
\widehat{\T}^*\|_{op}\, \|\widehat{\T}-\T\|_{op}\,\|\lambda(\T^*\T + \lambda \I)^{-1}\|_{op}
\,\|h_0\|_{\mathcal{H}}
\nonumber\\
& + \|\lambda (\widehat{\T}^*\widehat{\T}+\lambda \I)^{-1}\|_{op}
\,\|\widehat{\T}^*-\T^*\|_{op}\,
\|\T(\T^*\T+\lambda \I)^{-1}\|_{op} \,\|h_0\|_{\mathcal{H}} \\
=&
O_p\left(\frac{1}{\sqrt{n \lambda}}\right) \,
\end{align*}
uniformly in $\lambda\in\Lambda$.
Here we use that $\|\widehat{\T}-\T\|_{op}=
 \|\widehat{\T}^*-\T^*\|_{op} = O_p(1/\sqrt{n})$  from Lemma
 S3.2(f), and that   $\|\lambda(\T^*\T+\lambda
 \I)^{-1}\|_{op}\leq c$,
 $\|\lambda(\widehat{\T}^*\widehat{\T}+\lambda \I)^{-1}\|_{op}\leq c$,
 and  $\|(\widehat{\T}^*\widehat{\T} + \lambda \I )^{-1}
 \widehat{\T}^*\|_{op}\leq c'/\sqrt{\lambda}$   from Lemma S3.1. \\
 (e). From the proof of Proposition 3.11 in \cite{carrasco2007linear}
\begin{align*}
    \| (\T^* \T+\lambda \I)^{-1}\T^*
    \T h_0-h_0\|_{\mathcal{H}}^2\leq& \sup_j  \left[\frac{\lambda\sigma_j^\gamma}{\sigma_j^2+\lambda}\right]^2\sum_j\frac{|\left<h_0,\varphi_j\right>|^2}{\sigma_j^{2\gamma}}\,,\\
    \text{ with }\sup_j  &\left[\frac{\lambda\sigma_j^\gamma}{\sigma_j^2+\lambda}\right]^2\leq \max\left\{\left(\frac{\gamma}{2-\gamma}\right)^{\gamma},\sup_j \sigma_j^{2(\gamma-2)}\right\}\lambda^{2\wedge\gamma}\,,
\end{align*}
where we recall that $(\sigma_j,\varphi_j,\psi_j)_j$ is the singular system of the operator $\operatorname{T}$ (see Assumption 3.3). Now, by Assumption 3.3, $\sum_j|\left<h_0,\varphi_j\right>|^2\sigma_j^{2\gamma}<\infty$. Also,  since $\operatorname{T}$ is a compact operator, from Courant's Theorem, see \citet[Theorem 15.16]{kress1999linear}, we have that $\sigma_1=\|\operatorname{T}\|_{op}$, where $\sigma_1$ denotes the largest singular value of $\operatorname{T}$. So, by the above display we get (e).
    \hfill$\square$
}

\color{black}
\section{Convergence Rates and Hilbert Scales}
In this section, we obtain the convergence rate of our estimator as a function of the order of differentiability of $g_0$. This will require a link condition and an ``adaptivity" of our operator $\operatorname{T}$ to the space $\mathcal{H}$ (see the main paper for their definitions). We will use several concepts introduced in \cite{Carrasco2014} and \cite{engl1996regularization}. \\
Let $\operatorname{L}$ be an unbounded self-adjoint positive linear operator defined on a dense subset of the Hilbert space $\mathcal{E}=L^2[0,1]$. Let us denote with $\mathcal{D}(\operatorname{L}^k)$ the domain of $\operatorname{L}^k$ and let $\mathcal{M}=\cap_{k=0}^\infty \mathcal{D}(L^k)$.  We  let $\left<\cdot,\cdot\right>$ be the inner product on $L^2[0,1]$ and let $\|\cdot\|$ be the $L^2$ norm induced by it. For all $s\geq 0$ we introduce the inner product
\begin{equation*}
    \left<h,g\right>_s:=\left<L^s h, L^s g\right>
\end{equation*}
and we let $\|\cdot\|_s$ be the norm induced by $\left<\cdot,\cdot\right>_s$. Let $\mathcal{E}_s$ be the Hilbert space defined by the completion of $\mathcal{M}$ with respect to the norm $\|\cdot\|_s$. For $s\geq 0$ we have $\mathcal{E}_s=\mathcal{D}(L^s)$.
To give a specific form to the operator $\operatorname{L}$, let us introduce the integral operator
\begin{equation}
    (\operatorname{F}h)(z)=\int_{0}^z h(u) du\,,\, \operatorname{F}:L^2[0,1]\mapsto L^2[0,1]\, .
\end{equation}
The Hilbert adjoint $\operatorname{F}^*$ of $\operatorname{F}$ is given by
\begin{equation*}
    (\operatorname{F}^*g)(z)=\int_{z}^1 g(u) du\, .
\end{equation*}
In our case, we let $\operatorname{L}$ be such that
\begin{equation}\label{eq: definition of L minus 2}
\operatorname{L}^{-2}:=\operatorname{F}^* \operatorname{F}\, ,
\end{equation}
so that $\operatorname{L}$ will be just the inverse of the ``square root" of $\operatorname{F}^* \operatorname{F}$. This is an unbounded self-adjoint positive linear operator. From the definition of  $\operatorname{L}^{-2}$, we obtain that its inverse $\operatorname{L}^2$ is
\begin{equation*}
    \operatorname{L}^2 h=-h''\, .
\end{equation*}
We have that $\mathcal{D}(L^2)=\mathcal{H}$ , where $\mathcal{H}$ is defined in the main text of the paper.
\begin{remark}\label{rem: first derivative}
Notice that we cannot define $\operatorname{L} h=h'$, as the mapping $h\mapsto h'$ is not a self-adjoint operator. See \cite{Carrasco2014}.
\end{remark}
Recall that $\operatorname{T}=\operatorname{M}\operatorname{A}$, with $\operatorname{A}$ and $\operatorname{M}$ defined in the main text of the paper, 
  $g_0(z)=(1,z)\beta_0+h_0(z)$, and $h_0$ is the unique solution to the  integral equation
$\operatorname{M}\operatorname{E}\{Y \exp(\textbf{i}W^T \cdot)\}=\operatorname{T} h_0$\,.
To obtain a convergence rate for $\widehat g$ that depends on the degree of differentiability of $g_0$, we introduce the following ``adaptivity'' condition on the operator $\operatorname{T}$, also  called ``link condition'' in \cite{chen2011rate}:
\begin{assumption}\label{as: adaptivity} There exists a constant $a>0$ such that
    $\operatorname{T}$ satisfies
    \begin{equation*}
        \underline{m} \|h\|_{-a}\leq \|\operatorname{T}h\|\leq \overline{m}\|h\|_{-a}\,
    \end{equation*}
    for all $h\in \mathcal{E}$, for some fixed constants $\underline{m}<\overline{m}$.
\end{assumption}
The above assumption is similar to the ``link condition" and the ``reverse link condition" in \citet[Assumptions 2 and 5]{chen2011rate}, where {\itshape their} transformation $\phi$ is set as $\phi(u)=u^a$ and {\itshape their} unbounded operator $\operatorname{B}$ is equal to our unbounded operator $\operatorname{L}^{-1}$.
Notice that under the above assumption, the operator $\operatorname{T}$ is injective. The constant $a$ measures the degree of ill-posedness of the inverse problem, see \cite{Carrasco2014} and \cite{engl1996regularization}.
\begin{assumption}\label{as: differentiabiluty of h}
    $h_0 \in \mathcal{E}_b$ for some $b>0$\, .
\end{assumption}
When $b$ is an even number, the above condition ensures that $h_0$ is $b$ times differentiable.\\
We next obtain the convergence rate of $\widehat h$ as a function of $b$ (the degree of differentiability of $h_0$) and $a$ (the degree of ill-posed of the inverse problem). We will focus on the rate of convergence of $\widehat h$, as the convergence rate of $\widehat g(\cdot)=(1,\cdot)\widehat \beta +\widehat h(\cdot)$ is fully determined by rate of convergence of $\widehat h$. This is because, the rate of convergence of $\widehat \beta$ depends on $\widehat h$ (see the first equation in page 22 of the main paper). A result similar to the following can be found  in \citet[Proposition 3.1]{Carrasco2014}.
\begin{proposition}\label{prop: hilbert scale convergence rates}
    Let Assumptions 3.1 and 3.2 (from the main text) hold. Let $\lambda \rightarrow 0$ and $n \lambda \rightarrow \infty$. Also let Assumptions \ref{as: adaptivity} and \ref{as: differentiabiluty of h} hold and assume that $b\leq a+4$. Then, for $c=0,1, 2$ we have
    \begin{equation}
        \|\operatorname{L}^c \widehat h - \operatorname{L}^c h_0\|=O_P\left(\frac{1}{\sqrt{n \lambda}}+\lambda^{\frac{b-c}{2(a+2)}}\right)\, .
    \end{equation}
\end{proposition}
Before delving into the proof, let us comment on the result of the above proposition and the conditions at the basis of it. For $c\in\{0,2\}$, Proposition \ref{prop: hilbert scale convergence rates} obtains the convergence rate of the $L^2[0,1]$ norm of the structural function and its second derivative. For $c=1$, $\operatorname{L}\widehat h$ is {\itshape not exactly} the first derivative of $\widehat h$ (see Remark \ref{rem: first derivative}), but it can be interpreted as such. \\
The above result shows that (i) the smoother the function $h_0$ (i.e., the larger $b$), the quicker the convergence rate of our estimator, and (ii) the larger the ill-posedness of the inverse problem (i.e., the larger $a$), the slower the rate of convergence of our estimator.\\
Let us now clarify some aspects of the assumptions at the basis of Proposition \ref{prop: hilbert scale convergence rates}. Under Assumptions \ref{as: adaptivity} and \ref{as: differentiabiluty of h}, $h_0$ satisfies a ``source condition" involving the $L^2[0,1]$ inner product of $h_0$ and not of its second derivative. To see this, let $\operatorname{G}:=\operatorname{T}\operatorname{L}^{-2}$.
From \cite{engl1996regularization}, see also \cite{Carrasco2014}, when $|v|\leq 1$ we have that
\begin{equation}\label{as: source condition from hilbert scale}
    \mathcal{R}( (\operatorname{G}^*\operatorname{G})^{v/2})=\mathcal{E}_{v(a+2)}\, ,
\end{equation}
where $\mathcal{R}( (\operatorname{G}^*\operatorname{G})^{v/2})$ is the range of the operator $(\operatorname{G}^*\operatorname{G})^{v/2}$. 
 Also, $\mathcal{E}_s=\mathcal{D}(\operatorname{L}^s)$,
    as noticed earlier. Since $h_0\in\mathcal{E}_b$, see Assumption \ref{as: differentiabiluty of h}, for $v=b/(a+2)$ and $b\leq a+2$, Equation \eqref{as: source condition from hilbert scale} ensures that $h_0\in \mathcal{R}( (\operatorname{G}^*\operatorname{G})^{b/2(a+2)})$. 
    Thus, $h_0$ will still satisfy a source condition, but with respect to the (singular value decomposition of the) operator $\operatorname{G}$ and the inner product on $L^2[0,1]$. Accordingly, such a source condition involves the function $h_0$ and not its second derivative.\vspace{0.5cm}\\
    \textbf{Proof of Proposition \ref{prop: hilbert scale convergence rates}}. We divide the proof into three steps.\vspace{0.25cm}\\
    \textbf{Step 1}. Let   $s:=\operatorname{M}E[Y \exp(\textbf{i}W^T\cdot)]$ and  $\widehat s:=\widehat{\operatorname{M}} \,(1/n)\sum_{i=1}^n Y_i\exp(\textbf{i}W_i^T\cdot)$ its empirical counterpart.  We start by rewriting $\widehat h$ and $h_\lambda$ (defined in Equation (A.3)) in terms of the operators $\operatorname{L}^{-2}$, $\widehat{\operatorname{H}}:=\widehat{\operatorname{T}}\operatorname{L}^{-2}$, and $\operatorname{H}:=\operatorname{T}\operatorname{L}^{-2}$. As noticed earlier, $\mathcal{E}_2=\mathcal{D}(\operatorname{L}^2)=\mathcal{H}$. So, from the  the proof of Theorem 3.1, $\widehat h$ is the unique solution to the problem
\begin{align*}
\min_{h\in\mathcal{E}_2}\|\widehat{\operatorname{T}}h-\widehat s\|^2_\mu+\lambda\|h\|_{\mathcal{H}}^2=&\min_{h\in\mathcal{E}_2} \|\widehat{ \operatorname{T}}\operatorname{L}^{-2} \operatorname{L}^2 h- \widehat s\|^2_\mu+\lambda\|\operatorname{L}^2 h\|^2\\
=&\min_{h\in\mathcal{E}_2}\|\widehat{\operatorname{H}} \operatorname{L}^2 h - \widehat s\|^2_\mu + \lambda \|\operatorname{L}^2 h\|^2\, ,
\end{align*}
where $\|\cdot\|$ is the norm on $L^2[0,1]$, while $\|\cdot\|_\mu$ is the norm on $L^2_\mu$.  Now, $\widehat{\operatorname{H}}$ is a compact operator. To see this, notice first that $\operatorname{F}$ is a bounded operator, as $\|\operatorname{F}h\|^2=\int_{0}^1 \left|\int_{0}^1 1\{0\leq u\leq z\} h(u) du\right|^2 dz\leq \int_{0}^1 |h(u)|^2 du=\|h\|^2$. Thus, by \citet[Theorem 4.9]{kress1999linear}, also $\operatorname{F}^*$ (the Hilbert adjoint of $\operatorname{F}$)  is a bounded operator. Hence, $\operatorname{L}^{-2}=\operatorname{F}^* \operatorname{F}$ as a composition between two bounded operators will also be bounded.
From Lemma S3.2(e),  $\widehat{\operatorname{T}}$ is a compact operator. Hence, from \citet[Theorem 2.16]{kress1999linear}
$\widehat{\operatorname{H}}=\widehat{\operatorname{T}}\operatorname{L}^{-2}$ \,as a composition between a compact and a bounded operator is compact.
Since $\widehat{\operatorname{H}}$ is compact and hence bounded, we can apply \citet[Theorem 16.1]{kress1999linear} and get that $\operatorname{L}^2 \widehat h$ must satisfy\footnote{Arguing as in the proof of Theorem 3.1, we get that $(\widehat{\operatorname{H}}^*\widehat{\operatorname{H}} + \lambda \operatorname{I})$ is a coercive operator, and hence it has a bounded inverse.}
\begin{equation}\label{eq: reformulation of hhat lambda}
    \operatorname{L}^2 \widehat h=(\widehat{\operatorname{H}}^* \widehat{\operatorname{H}} + \lambda \operatorname{I})^{-1}\widehat{\operatorname{H}}^* \widehat s\, \Leftrightarrow  \widehat h=\operatorname{L}^{-2}(\widehat{\operatorname{H}}^* \widehat{\operatorname{H}} + \lambda \operatorname{I})^{-1}\widehat{\operatorname{H}}^* \widehat s\, .
\end{equation}
By similar arguments, since $h_\lambda$  solves $\min_{h\in\mathcal{E}_2}\|\operatorname{T} h - s\|^2_\mu+\lambda \|h\|^2$, it must satisfy
\begin{equation}\label{eq: reformulation of h lambda}
        \operatorname{L}^2  h_\lambda=(\operatorname{H}^* \operatorname{H} + \lambda \operatorname{I})^{-1}\operatorname{H}^*  s\, \Leftrightarrow  h_\lambda=\operatorname{L}^{-2}(\operatorname{H}^* \operatorname{H} + \lambda \operatorname{I})^{-1}\operatorname{H}^*  s\, ,
\end{equation}
where $\operatorname{H}$ is a compact operator.\vspace{0.25cm}\\
\textbf{Step 2}. Next, 
for $c=0,1,2$ ,
\begin{equation}\label{eq: bias variance decomposition of hhat lambda}
    \|\operatorname{L}^c \widehat h - \operatorname{L}^c h_0\|\leq \|\operatorname{L}^c \widehat h - \operatorname{L}^c  h_\lambda\|+\|\operatorname{L}^c h_\lambda - \operatorname{L}^c h_0\|\,.
\end{equation}
The first term on the right-hand side represents a ``variance term", while the second term is a ``(regularization) bias term". Let us deal with the variance term first. Notice  that as long as $\operatorname{L}^{c-2}$ is a bounded operator, we have
\begin{equation}\label{eq: inequality for the variance term}
    \|\operatorname{L}^c \widehat h - \operatorname{L}^c  h_\lambda\|=\|\operatorname{L}^{c-2}(\operatorname{L}^2 \widehat h - \operatorname{L}^2  h_\lambda)\|\leq \|\operatorname{L}^{c-2}\|_{op}\,\|\operatorname{L}^2 \widehat h - \operatorname{L}^2  h_\lambda\| \, .
\end{equation}
To see that $\|\operatorname{L}^{c-2}\|_{op}<\infty$ for $c=0,1,2$, notice first that for $c=2$ we trivially have $\operatorname{L}^{c-2}=\operatorname{I}$ (the identity operator). For $c=0$, $\operatorname{L}^{c-2}=\operatorname{L}^{-2}$ that is bounded, as noticed earlier. Consider now the case $c=1$, so $\operatorname{L}^{c-2}=\operatorname{L}^{-1}=(\operatorname{L}^{-2})^{1/2}=(\operatorname{F}^* \operatorname{F})^{1/2}$. Since $\operatorname{F}h=\int_{0}^1 1\{0\leq u\leq z\} h(u) du$ with $\int_{0}^1 \int_{0}^1 |1(0\leq u\leq z)|^2 du dz<\infty$, $\operatorname{F}$ is an integral operator whose kernel is Hilbert-Schmidt (square integrable), so by \citet[Proposition 2.1]{Busby1972} $\operatorname{F}$ is compact. As a compact operator, it admits a singular system, see \citet[Theorem 15.16]{kress1999linear}. Let $(\mu_j,\phi_j, \eta_j)_j$ be the singular system of $\operatorname{F}$, where the singular values $(\mu_j)_j$ are ordered in a decreasing order, $(\phi_j)_j$ is a sequence of orthonormal elements of $L^2[0,1]$, and $(\eta_j)_j$ is another sequence of orthonormal elements in $L^2[0,1]$. Then, $(\operatorname{F}^* \operatorname{F})^{1/2}h=\sum_j \mu_j \left<h,\phi_j\right>\phi_j$, see \cite{florens2011identification}. From \citet[Theorem 15.17]{kress1999linear} the largest singular value satisfies $\mu_1\leq \|\operatorname{F}\|_{op}<\infty$. We therefore have that $\|(\operatorname{F}^*\operatorname{F})^{1/2}h\|^2=\|\sum_j \mu_j \left<h,\phi_j\right>\phi_j\|^2=\sum_j \mu_j^2 |\left<h,\phi_j\right>|^2$ $\leq \|F\|^2_{op} \sum_j |\left<h,\phi_j\right>|^2$ $\leq \|F\|^2_{op} \|h\|^2$, where the last inequality follows from Bessel's inequality. This shows that $(\operatorname{F}^*\operatorname{F})^{1/2}$, hence $\operatorname{L}^{-1}$, is a bounded operator. Thus, we obtain that $\|\operatorname{L}^{c-2}\|_{op}<\infty$ for $c=0,1,2$ in Equation \eqref{eq: inequality for the variance term}.\\
Next, to get a rate for  $\|\operatorname{L}^2\widehat h - \operatorname{L}^2 h_\lambda\|$, we use Equations (\ref{eq: reformulation of hhat lambda}) and (\ref{eq: reformulation of h lambda}) and get
\begin{equation*}
    \|\operatorname{L}^2\widehat h - \operatorname{L}^2 h_\lambda\|=\|(\widehat{\operatorname{H}}^* \widehat{\operatorname{H}} + \lambda \operatorname{I})^{-1}\widehat{\operatorname{H}}^* \widehat s - (\operatorname{H}^* \operatorname{H} + \lambda \operatorname{I})^{-1}\operatorname{H}^*  s\|\, .
\end{equation*}
Notice that
\begin{align*}
\widehat{\operatorname{H}}h=&\widehat{\operatorname{T}}\operatorname{L}^{-2} h=\widehat{\operatorname{T}}\operatorname{F}^* \operatorname{F} h=\widehat{\operatorname{M}}\widehat{\operatorname{A}}\operatorname{F}^* \operatorname{F} h\\
=&\widehat{\operatorname{M}} \int_{0}^1 \left[(1/n)\sum_{i=1}^n \exp(W_i^T\cdot)\int_{0}^1 1(Z_i\vee u \leq v \leq 1) dv\right] h(u) du\, \\
\text{ and } \operatorname{H}h=&\operatorname{M}\int_{0}^1 \left[E \exp(W^T\cdot)\int_{0}^1 1(Z\vee u \leq v \leq 1) dv\right] h(u) du\, .
\end{align*}
By the above expressions and by using arguments similar to those used in proof of Lemma S 3.2 (d)(f), we obtain that
\begin{equation*}
    \|\widehat{\operatorname{H}}-\operatorname{H}\|_{op}=O_P\left(\frac{1}{\sqrt{n}}\right)\, .
\end{equation*}
Thus, arguments similar to those used in the proof of Theorem 3.1 lead to
\begin{equation*}
    \|(\widehat{\operatorname{H}} \widehat{\operatorname{H}}^* + \lambda \operatorname{I})^{-1}\widehat{\operatorname{H}}^* \widehat s - (\operatorname{H} \operatorname{H}^* + \lambda \operatorname{I})^{-1}\operatorname{H}^* s\|=O_P\left(\frac{1}{\sqrt{n \lambda}}\right)\, .
\end{equation*}
Gathering results we obtain the following rate for the ``variance term" (with $c=0,1,2$)
\begin{equation}\label{eq: rate for the variance term}
    \|\operatorname{L}^c \widehat h - \operatorname{L}^c h_\lambda\|=O_P\left(\frac{1}{\sqrt{n \lambda}}\right)\, .
\end{equation}
\textbf{Step 3}. We now get a convergence rate for the ``bias term" in (\ref{eq: bias variance decomposition of hhat lambda}). We have (see the comments below)
\begin{align}\label{eq: variance reformulation 1}
    \|\operatorname{L}^c h_\lambda - \operatorname{L}^c h_0\|=&\|\operatorname{L}^c \operatorname{L}^{-2}(\operatorname{H}^* \operatorname{H}+\lambda \operatorname{I})^{-1}\operatorname{H}^* s - \operatorname{L}^{c-2}\operatorname{L}^2 h_0\|\nonumber\\
    =&\|\operatorname{L}^{c-2}[(\operatorname{H}^* \operatorname{H}+\lambda \operatorname{I})^{-1}\operatorname{H}^* s - \operatorname{L}^2 h_0]\|\nonumber\\
    =& \|(\operatorname{H}^* \operatorname{H}+\lambda \operatorname{I})^{-1}\operatorname{H}^* s - \operatorname{L}^2 h_0\|_{c-2}\nonumber\\
    =& \|(\operatorname{H}^* \operatorname{H}+\lambda \operatorname{I})^{-1}\operatorname{H}^* \operatorname{T}\operatorname{L}^{-2}\operatorname{L}^2 h_0 -\operatorname{L}^2 h_0 \|_{c-2}\nonumber\\
    =& \|[(\operatorname{H}^* \operatorname{H}+\lambda \operatorname{I})^{-1}\operatorname{H}^*\operatorname{H} - \operatorname{I}] \operatorname{L}^2 h_0\|_{c-2}\,,
\end{align}
where the first equality follows from from Equation (\ref{eq: reformulation of h lambda}), the third equality follows from the definition of the norm $\|\cdot\|_{s}$ (for $s>0$), the fourth equality follows from the fact that $\operatorname{T}h_0=s$ (see the proof of Theorem 3.1), while the last inequality follows from $\operatorname{H}=\operatorname{T}\operatorname{L}^{-2}$.  \\
Now, from Assumption \ref{as: adaptivity} and \citet[Corollary 8.22]{engl1996regularization}, $\underline{c}(v)\|h\|_{-v(a+2)}\leq \|(\operatorname{H}^* \operatorname{H})^{v/2}h\|\leq \overline{c}(v)\|h\|_{-v(a+2)}$ for $|v|\leq 1$ and fixed constants $\underline{c}(v)$ and $\overline{c}(v)$ . By applying the first inequality with $-v=(c-2)/(a+2)$ (for $c\in\{0,1,2\}$) and $h=[(\operatorname{H}^*\operatorname{H}+\lambda \operatorname{I})^{-1}\operatorname{H}^*\operatorname{H}- \operatorname{I}]\operatorname{L^2 h_0}$, we get that the right-hand side of the previous display can be bounded as
\begin{equation}\label{eq: variance bound}
    \|[(\operatorname{H}^*\operatorname{H}+\lambda \operatorname{I})^{-1}\operatorname{H}^*\operatorname{H}- \operatorname{I}]\operatorname{L^2 h_0}\|_{c-2}\leq \frac{1}{\underline{c}(v)}\|(\operatorname{H}^* \operatorname{H})^{\frac{2-c}{2(a+2)}}[(\operatorname{H}^*\operatorname{H}+\lambda \operatorname{I})^{-1}\operatorname{H}^*\operatorname{H}- \operatorname{I}]\operatorname{L^2 h_0}\|
\end{equation}
Next, by Assumption \ref{as: differentiabiluty of h} $h_0\in\mathcal{E}_b$ and, as noticed earlier,  $\mathcal{E}_s=\mathcal{D}(\operatorname{L}^s)$ for any $s>0$. So, $\operatorname{L}^b h_0$ is a well defined element of $L^2[0,1]$ and so is $\operatorname{L}^{b-2} (\operatorname{L}^{2}h_0)$. Thus,  $\operatorname{L}^2 h_0\in\mathcal{D}(\operatorname{L}^{b-2})=\mathcal{E}_{b-2}$. Next,  \citet[Corollary 8.22]{engl1996regularization} ensures that $\mathcal{R}((\operatorname{H}^*\operatorname{H})^{v/2})=\mathcal{E}_{v(a+2)}$ for $|v|\leq 1$, where $\mathcal{R}((\operatorname{H}^*\operatorname{H})^{v/2})$ is the range of $(\operatorname{H}^*\operatorname{H})^{v/2}$. Using the latter equality with  $v=(b-2)/(a+2)$ we get that $\mathcal{R}((\operatorname{H}^*\operatorname{H})^{\frac{b-2}{2(a+2)}})=\mathcal{E}_{b-2}$, so that $\operatorname{L}^2 h_0\in \mathcal{R}((\operatorname{H}^*\operatorname{H})^{\frac{b-2}{2(a+2)}})$. This implies that there exists $\eta\in L^2[0,1]$ such that
$
(\operatorname{H}^*\operatorname{H})^{\frac{b-2}{2(a+2)}}\eta=\operatorname{L}^2 h_0\,$. So,
\begin{align}\label{eq: variance reformulation 2}
   \|(\operatorname{H}^* \operatorname{H})^{\frac{2-c}{2(a+2)}}[(\operatorname{H}^*\operatorname{H}+\lambda \operatorname{I})^{-1}&\operatorname{H}^*\operatorname{H}- \operatorname{I}]\operatorname{L^2 h_0}\|\nonumber\\
   =& \|(\operatorname{H}^* \operatorname{H})^{\frac{2-c}{2(a+2)}}[(\operatorname{H}^*\operatorname{H}+\lambda \operatorname{I})^{-1}\operatorname{H}^*\operatorname{H}- \operatorname{I}] (\operatorname{H}^*\operatorname{H})^{\frac{b-2}{2(a+2)}}\eta\|\nonumber\\
    =&\|(\operatorname{H}^*\operatorname{H})^{\frac{2-c}{2(a+2)}}(\operatorname{H}^*\operatorname{H}+\lambda \operatorname{I})^{-1}(\operatorname{H}^*\operatorname{H})^{1+\frac{b-2}{2(a+2)}}\eta - (\operatorname{H}^*\operatorname{H})^{\frac{b-c}{2(a+2)}}\eta\|\, .
\end{align}
Let us now deal with the difference on the RHS of the previous display. As noticed earlier, $\operatorname{H}$ is a compact operator. Hence, from \citet[Theorem 15.16]{kress1999linear} it admits a singular system that we denote as $(\widetilde \sigma_j,\widetilde \phi_j,\widetilde \psi_j)_j$, where $(\widetilde \sigma_j)_j$ is a sequence of singular values, $(\widetilde \phi_j)_j$ is a sequence of orthonormal elements in $L^2[0,1]$, and $(\widetilde \psi_j)_j$ is another sequence of orthonormal elements in $L^2[0,1]$. From \cite{florens2011identification} we have that $(\operatorname{H}^*\operatorname{H})^\alpha h=\sum_j\widetilde \sigma_j^{2\alpha} \left<h,\widetilde \phi_j\right> \widetilde \phi_j$ for any $\alpha>0$. Hence,
\begin{equation*}
(\operatorname{H}^*\operatorname{H})^{1+\frac{b-2}{2(a+2)}}\eta=\sum_j \widetilde \sigma_j^{2+\frac{b-2}{a+2}} \left<\eta,\widetilde \phi_j\right> \widetilde \phi_j\, .
\end{equation*}
Also, \citet[Theorem 15.16]{kress1999linear} ensures that $\operatorname{H}\widetilde \phi_j=\widetilde \sigma_j \widetilde \psi_j$ and $\operatorname{H}^*\widetilde \psi_j=\widetilde \sigma_j \widetilde \phi_j$. Thus, $\operatorname{H}^*\operatorname{H}\widetilde\phi_j=\widetilde\sigma_j^2\widetilde\phi_j$ and $(\operatorname{H}^*\operatorname{H}+\lambda \operatorname{I})\widetilde\phi_j=(\widetilde\sigma_j^2+\lambda)\widetilde\phi_j$. This implies that $(\operatorname{H}^*\operatorname{H}+\lambda \operatorname{I})^{-1}\widetilde\phi_j=\widetilde\phi_j/(\widetilde\sigma_j^2+\lambda)$. Using this and the previous display leads to
\begin{align*}
(\operatorname{H}^*\operatorname{H}+\lambda \operatorname{I})^{-1} (\operatorname{H}^*\operatorname{H})^{1+\frac{b-2}{2(a+2)}}\eta=&(\operatorname{H}^*\operatorname{H}+\lambda \operatorname{I})^{-1}\sum_j \widetilde \sigma_j^{2+\frac{b-2}{a+2}} \left<\eta,\widetilde \phi_j\right> \widetilde \phi_j\\
=&\sum_j \widetilde \sigma_j^{2+\frac{b-2}{a+2}} \left<\eta,\widetilde \phi_j\right> (\operatorname{H}^*\operatorname{H}+\lambda \operatorname{I})^{-1}\widetilde \phi_j\\
=& \sum_j \widetilde \sigma_j^{2+\frac{b-2}{a+2}} \left<\eta,\widetilde \phi_j\right> \frac{1}{\widetilde \sigma_j^2+\lambda}\widetilde \phi_j\,,
\end{align*}
where in the second equality we have used the fact that $(\operatorname{H}^*\operatorname{H}+\lambda \operatorname{I})^{-1}$ is a bounded operator for any $\lambda>0$, so it can be exchanged with $\sum_j$. Since $(\operatorname{H}^*\operatorname{H})^\alpha h=\sum_j\widetilde \sigma_j^{2\alpha} \left<h,\widetilde \phi_j\right> \widetilde \phi_j$ for any $\alpha>0$, by the orthonormality of $(\widetilde \phi_j)_j$ we get that $(\operatorname{H}^*\operatorname{H})^{\frac{2-c}{2(a+2)}}\widetilde \phi_j=\widetilde \sigma_j^{\frac{2-c}{a+2}}\widetilde \phi_j$. By this last equality and the previous display, we obtain that
\begin{align*}
(\operatorname{H}^*\operatorname{H})^{\frac{2-c}{2(a+2)}}(\operatorname{H}^*\operatorname{H}+\lambda \operatorname{I})^{-1} (\operatorname{H}^*\operatorname{H})^{1+\frac{b-2}{2(a+2)}}\eta=&(\operatorname{H}^*\operatorname{H})^{\frac{2-c}{2(a+2)}}\sum_j \widetilde \sigma_j^{2+\frac{b-2}{a+2}} \left<\eta,\widetilde \phi_j\right> \frac{1}{\widetilde \sigma_j^2+\lambda}\widetilde \phi_j\\
=& \sum_j \widetilde \sigma_j^{2+\frac{b-2}{a+2}} \left<\eta,\widetilde \phi_j\right> \frac{1}{\widetilde \sigma_j^2+\lambda}(\operatorname{H}^*\operatorname{H})^{\frac{2-c}{2(a+2)}}\widetilde \phi_j\\
=& \sum_j \widetilde \sigma_j^{\frac{b-c}{a+2}} \left<\eta,\widetilde \phi_j\right> \frac{\widetilde \sigma_j^2}{\widetilde \sigma_j^2+\lambda}\widetilde \phi_j\, .
\end{align*}
By using the previous display and the fact that $(\operatorname{H}^*\operatorname{H})^{\frac{b-c}{2(a+2)}}\eta=\sum_j \widetilde\sigma_j^{\frac{b-c}{a+2}}\left<\eta,\widetilde\phi_j\right>\widetilde\phi_j$, as  noticed earlier, we obtain that the square of the norm on the RHS of (\ref{eq: variance reformulation 2}) can be written as
\begin{align}\label{eq: bound for the variance term}
    \|(\operatorname{H}^*\operatorname{H})^{\frac{2-c}{2(a+2)}}(\operatorname{H}^*\operatorname{H}+\lambda \operatorname{I})^{-1}&(\operatorname{H}^*\operatorname{H})^{1+\frac{b-2}{2(a+2)}}\eta - (\operatorname{H}^*\operatorname{H})^{\frac{b-c}{2(a+2)}}\eta\|^2\nonumber\\
    =& \|\sum_j \widetilde \sigma_j^{\frac{b-c}{a+2}} \left<\eta,\widetilde \phi_j\right> \frac{\widetilde \sigma_j^2}{\widetilde \sigma_j^2+\lambda}\widetilde \phi_j - \sum_j \widetilde\sigma_j^{\frac{b-c}{a+2}}\left<\eta,\widetilde\phi_j\right>\widetilde\phi_j\|^2\nonumber\\
    =& \|\sum_j \widetilde\sigma_j^{\frac{b-c}{a+2}}\left(\frac{\widetilde \sigma_j^2}{\widetilde \sigma_j^2+\lambda}-1\right)\left<\eta,\widetilde\phi_j\right>\widetilde\phi_j\|^2\nonumber\\
    =& \sum_j \left[\widetilde\sigma_j^{\frac{b-c}{a+2}}\left(\frac{\widetilde\sigma_j^2}{\widetilde\sigma_j^2+\lambda}-1\right)\right]^2|\left<\eta,\widetilde\phi_j\right>|^2\nonumber\\
    \leq & \max_{\sigma>0}\left[\sigma^{\frac{b-c}{2(a+2)}}\left(\frac{\sigma}{\sigma+\lambda}-1\right)\right]^2 \sum_j |\left<\eta,\widetilde\phi_j\right>|^2\nonumber\\
    \leq & \lambda^{\frac{b-c}{a+2}}|\sum_j\left<\eta,\widetilde\phi_j\right>|^2\nonumber\\
    \leq & \lambda^{\frac{b-c}{a+2}} \|\eta\|^2\,,
\end{align}
where in the last two inequalities we have used $\max_{\sigma>0}\left[\sigma^{\frac{b-c}{2(a+2)}}\left(\frac{\sigma}{\sigma+\lambda}-1\right)\right]^2\leq \lambda^{\frac{b-c}{a+2}}$ and $\sum_j\left<\eta,\widetilde\phi_j\right>|^2\leq \|\eta\|$ (from Bessel's inequality). \\
Finally, gathering Equations (\ref{eq: bound for the variance term}), (\ref{eq: variance reformulation 2}), (\ref{eq: variance bound}), and (\ref{eq: variance reformulation 1}) gives the convergence rate for the bias term,
\begin{equation}\label{eq: rate for bias term}
    \|\operatorname{L}^c h_\lambda - \operatorname{L}^c h_0\|=O(\lambda^{\frac{b-c}{2(a+2)}})\, .
\end{equation}
From (\ref{eq: rate for bias term}), (\ref{eq: rate for the variance term}), and (\ref{eq: bias variance decomposition of hhat lambda}), we obtain the desired result.
   \hfill$\square$

\section{Analysis of cross-validation}
We consider random sample splitting, where an estimator
$\widehat{g}_\lambda$ is obtained from the first  fold of size $n_1 =
\lfloor n/2\rfloor$ (the largest integer smaller than $n/2$), and the second  fold of size $n_2= n -
n_1$ is used for validation. Then
\begin{equation}
M_{n_2}(\widehat{g}_\lambda)
 = \int \left| \frac{1}{n_2} \sum_{i=n_1 +1}^{n}{ \left( Y_i -
 \widehat{g}_\lambda (Z_i) \right)
\exp(\mathbf{i}W_i^\top t) } \right|^2 \, d\mu(t)
\nonumber
\label{Mnlambda}
\end{equation}
is minimized over a  set
 $\Lambda=[\underline{\lambda}_n,\overline{\lambda}_n]\subset
 \R_+$ of  penalty parameters. For notational simplicity, the following analysis will  focus on the above objective function, but it can be straightforwardly extended to a k-fold cross-validation method.
\begin{proposition}
Let Assumptions 3.1, 3.2, and 3.3  hold. Assume that $n \underline{\lambda}_n \rightarrow \infty$,
$\overline{\lambda}_n\rightarrow 0$, 
then
$M_{n_2}(\widehat{g}_\lambda) =  O_P \left(\frac{1}{{n\lambda}} +
\lambda^{\gamma \wedge 2}\right)$ uniformly in $\lambda \in \Lambda$.
\end{proposition}

\begin{proof}
Consider the decomposition
\begin{align*}
\frac{1}{n_2} \sum_{i=n_1+1}^{n}{ \left( Y_i - \widehat{g}_\lambda (Z_i) \right)
\exp(\mathbf{i}W_i^\top t) }&  =
\frac{1}{n_2} \sum_{i=n_1+1}^{n}{ \left( Y_i - {g}_0 (Z_i) \right)
\exp(\mathbf{i}W_i^\top t) }
\\ \mbox{}  & +
\frac{1}{n_2} \sum_{i=n_1+1}^{n}{ \left( {g}_0 (Z_i) - \widehat{g}_\lambda (Z_i) \right)
\exp(\mathbf{i}W_i^\top t) }
\, .
\end{align*}

For the first term, we show that
\begin{equation}\label{eq: first term of CV decomposition}
\left\|\frac{1}{n_2}\sum_{i=n_1+1}^n (Y_i-g_0(Z_i))\exp(\textbf{i}W_i^T\cdot)\right\|_\mu=O_P(n_2^{-1/2})\, ,
\end{equation}
where $\|\cdot\|_\mu$ denotes the norm on $L^2_\mu(\mathcal{T})$, the space of functions defined on $\mathcal{T}$ that are square integrable with respect to $\mu$. We get (see the comments below)
\begin{align*}
\operatorname{E}\Big\|\frac{1}{\sqrt{n_2}}\sum_{i=n_1+1}^n (Y_i-g_0(Z_i))&\exp(\textbf{i}W_i^T\cdot)\Big\|_\mu^2\\
&=\operatorname{E}\left[\frac{1}{n_2}\sum_{i,j=n_1+1}^n (Y_i - g_0(Z_i)) \mathcal{F}_\mu(W_i-W_j) (Y_j-g_0(Z_j)) \right]\\
&=\operatorname{E}\left[\frac{1}{n_2}\sum_{i=n_1+1}^n (Y_i - g_0(Z_i))^2 \mathcal{F}_\mu(0)^2\right]\\
&= \mathcal{F}_\mu(0)^2 \operatorname{E}[(Y_i- g_0(Z_i))^2]\, .
\end{align*}
The first equality simply follows from developing the square norm $\|\cdot\|_\mu^2$ and the fact that $\mathcal{F}_\mu$ is the characteristic function of $\mu$. To get the second equality, notice that the cross-terms with $i\neq j$ cancel by the iid structure of the data and $\operatorname{E}[Y_i-g_0(Z_i)|W_i]=0$. Since $\operatorname{E}[(Y_i-g_0(Z_i))^2]<\infty$,  by the previous display we get \eqref{eq: first term of CV decomposition}.\\

For the second term of the decomposition,
\[
\left| \frac{1}{n_2} \sum_{i=n_1+1}^{n}{ \left( {g}_0 (Z_i) -
\widehat{g}_\lambda (Z_i) \right)
\exp(\mathbf{i}W_i^\top t)} \right|
\leq
\sup_z |g_0(z) - \widehat{g}_\lambda(z)|
\, .
\]
Thus, as long as
\begin{equation}\label{equation: uniform in lambda rate of convergence}
    \sup_z |g_0(z) - \widehat{g}_\lambda(z)|
= O_P \left( \frac{1}{\sqrt{n\lambda}} + \lambda^{\frac{\gamma \wedge 2}{2}}\right)
\end{equation}
uniformly in $\lambda  \in \Lambda$, we get that
\[
M_{n_2}(\widehat{g}_\lambda)  =
\int \left|
\frac{1}{n_2} \sum_{i=n_1+1}^{n}{ \left( Y_i - \widehat{g}_\lambda (Z_i) \right)
\exp(\mathbf{i}W_i^\top t) } \right|^2 \, d\mu(t)
= O_P\left( n^{-1} +
\frac{1}{{n\lambda}} + \lambda^{\gamma \wedge 2} \right)
\, ,
\]
uniformly in $\lambda \in \Lambda$.
Use that $n \underline{\lambda}_n\rightarrow \infty$  and $\overline{\lambda}_n\rightarrow 0$ to conclude. \\

To show Equation \eqref{equation: uniform in lambda rate of convergence}, let us recall that $\widehat g_\lambda(z)=(1,z)\widehat \beta_\lambda+\widehat h_\lambda(z)$ and $g_0(z)=(1,z)\beta_0+h_0(z)$, where $\widehat h_\lambda$ and $\widehat \beta_\lambda$ are defined as in Equations (A.9) and (A.10), while $\beta_0$ and $h_0$ are as in Step 1 of the proof of Theorem 3.1. Using that $\|\widehat h_\lambda\|\leq \|h_0\|+\|\widehat h_\lambda-h_0\|$ and arguments similar to those in Step 4 of the proof of  Theorem 3.1 lead to
 \begin{equation}\label{eq: betahat intermediate result uniform in lambda}
     \|\widehat \beta_\lambda - \beta_0\|=O_P\left(n^{-1/2}+\|\widehat h_\lambda-h_0\|_{\mathcal{H}}\right)\,\text{ uniformly in } \lambda\in\Lambda\, .
 \end{equation}
 Next, $\widehat h_\lambda - h_0=S_{1,\lambda}+S_{2,\lambda}+S_{3,\lambda}+S_{4,\lambda}+h_\lambda-h_0$, where
\begin{gather*}
    S_1=(\T^* \T + \lambda \I )^{-1}\T^* (\widehat{\M}\widehat r-
    \widehat{\T} h_0)\, , \quad
    S_2=(\T^* \T + \lambda \I )^{-1}(\widehat{\T}^*-
    \T^*)(\widehat{\M}\widehat r - \widehat{\T}h_0)\, ,\\
    S_3= \left[(\widehat{\T}^* \widehat{\T}+\lambda \I)^{-1} - (\T^* \T
    + \lambda \I )^{-1} \right]\widehat{\T}^* (\widehat{\M}\widehat r
    - \widehat{\T}h_0)\, ,  \quad
    S_4=(\widehat{\T}^* \widehat{\T}+\lambda \I)^{-1}\widehat{\T}^*
    \widehat{\T}h_0 - h_\lambda\,, \\
    h_\lambda=(\operatorname{T}^*\operatorname{T}+\lambda \operatorname{I})^{-1}\operatorname{T}^*\operatorname{T}h_0\, .
\end{gather*}
A direct application of Lemma S3.3 gives
\begin{gather*}
    \|S_{j,\lambda}\|_{\mathcal{H}}=O_P\left(\frac{1}{\sqrt{n\lambda}}\right)\text{ uniformly in }\lambda\in\Lambda\, \text{ for }j=1,2,3,4
\end{gather*}
and
\begin{equation*}
    \|h_\lambda-h_0\|_{\mathcal{H}}=O_P(\lambda^{(\gamma\wedge2)/2})\text{ uniformly in }\lambda\in\Lambda\, .
\end{equation*}
Thus,
\begin{equation}\label{eq: eq: hhat intermediate result uniform in lambda}
    \|\widehat h_\lambda - h_0\|_{\mathcal{H}}=O_P\left(\frac{1}{\sqrt{n \lambda}}+\lambda^{(\gamma\wedge2)/2}\right)\text{ uniformly in }\lambda\in\Lambda\, .
\end{equation}
By \eqref{eq: eq: hhat intermediate result uniform in lambda} and \eqref{eq: betahat intermediate result uniform in lambda},
\begin{align*}
    \|\widehat \beta_\lambda-\beta_0\|=&O_P\left(n^{-1/2}+\frac{1}{\sqrt{n \lambda}}+\lambda^{(\gamma\wedge2)/2}\right)\\
    =&O_P\left(\frac{1}{\sqrt{n \lambda}}+\lambda^{(\gamma\wedge2)/2}\right)\text{ uniformly in }\lambda\in\Lambda\, ,
\end{align*}
where the last equality follows from $\overline{\lambda}_n\rightarrow 0$.
As noticed in Step 4 of the proof of Theorem 3.1, $\sup_{z\in[0,1]}|\widehat h_\lambda(z)-h_0(z)|\leq \|\widehat h_\lambda-h_0\|_\mathcal{H}$, so by \eqref{eq: eq: hhat intermediate result uniform in lambda}
\begin{equation*}
  \sup_{z\in[0,1]}|\widehat h_\lambda(z)-h_0(z)|=O_P\left(\frac{1}{\sqrt{n \lambda}}+\lambda^{(\gamma\wedge2)/2}\right)\text{ uniformly in }\lambda\in\Lambda\, .
\end{equation*}
As recalled earlier, $\widehat g_\lambda(z)=(1,z)\widehat \beta_\lambda+\widehat h_\lambda(z)$ and $g_0(z)=(1,z)\beta_0+h_0(z)$, so the previous two displays give (\ref{equation: uniform in lambda rate of convergence}) and conclude the proof.

\end{proof}

\end{document}